\documentclass[a4paper,12pt]{article}

\usepackage{amsfonts}
\usepackage{mathrsfs}
\usepackage{amsmath}
\usepackage{amssymb}
\usepackage{framed}

\usepackage[medium]{titlesec}
\usepackage{bm}
\usepackage{cite}
\usepackage{stmaryrd}

\usepackage{cancel}

\usepackage[normalem]{ulem}
\usepackage{extarrows}
\usepackage{slashed}
\usepackage{isodateo}
\usepackage{graphicx}
\usepackage[dvipsnames]{xcolor}
\usepackage[bookmarksnumbered=true,bookmarksopen=true]{hyperref}
 \hypersetup{colorlinks,%
             linkcolor=NavyBlue, %
             citecolor=PineGreen, %
             urlcolor=PineGreen}
\usepackage[hmargin=.7in,vmargin=1.1in]{geometry}
\usepackage{indentfirst}
\usepackage{booktabs}
\usepackage{multirow}

\usepackage{bbm}

\newcommand{\FR}[2]{\displaystyle\frac{\,{#1}\,}{#2}}
\newcommand{\fr}[2]{\mbox{$\frac{\,{#1}\,}{#2}$}}
\newcommand{\n}{\nonumber}

\makeatletter
\newcommand{\subalign}[1]{%
  \vcenter{%
    \Let@ \restore@math@cr \default@tag
    \baselineskip\fontdimen10 \scriptfont\tw@
    \advance\baselineskip\fontdimen12 \scriptfont\tw@
    \lineskip\thr@@\fontdimen8 \scriptfont\thr@@
    \lineskiplimit\lineskip
    \ialign{\hfil$\m@th\scriptstyle##$&$\m@th\scriptstyle{}##$\hfil\crcr
      #1\crcr
    }%
  }%
}
\makeatother

\graphicspath{{fig/}}

\def\bge{\begin{equation}}
\def\ede{\end{equation}}
\def\bga{\begin{aligned}}
\def\eda{\end{aligned}}
\def\bgb{\begin{bmatrix}}
\def\edb{\end{bmatrix}}
\def\bgp{\begin{pmatrix}}
\def\edp{\end{pmatrix}}
\def\bgm{\begin{matrix}}
\def\edm{\end{matrix}}
\def\bgs{\begin{subequations}}
\def\eds{\end{subequations}}
\newcommand{\order}[1]{\mathcal{O}({#1})}
\def\di{{\mathrm{d}}}

\def\pd{\partial}
\def\ld{{\mathscr{L}}}

\def\la{\langle}\def\ra{\rangle}

\def\to{\rightarrow}

\def\ii{\mathrm{i}}

\def\al{\alpha}
\def\be{\beta}

\def\de{\delta}
\def\ep{\epsilon}

\def\lam{\lambda}

\def\si{\sigma}

\def\aa{\mathsf{a}}

\def\cc{\mathsf{c}}
\def\dd{\mathsf{d}}

\def\Re{\mathrm{Re}\,}
\def\Im{\mathrm{Im}\,}

\def\G{{\mathcal{G}}}

\usepackage{mdframed} 
\usepackage{tikz}

\newmdenv[skipabove=0pt,%
          skipbelow=5pt,%
          leftmargin=0pt,%
          rightmargin=0pt,%
          innertopmargin=-5pt,%
          innerbottommargin=7pt,%
          innerleftmargin=2pt,%
          innerrightmargin=2pt,%
          splittopskip=0pt,%
          splitbottomskip=0pt,%
          linewidth=0pt,%
          nobreak=true]%
          {keyeqn2}

\newmdenv[backgroundcolor=gray!15,%
          skipabove=0pt,%
          skipbelow=5pt,%
          leftmargin=0pt,%
          rightmargin=0pt,%
          innertopmargin=-5pt,%
          innerbottommargin=7pt,%
          innerleftmargin=2pt,%
          innerrightmargin=2pt,%
          splittopskip=0pt,%
          splitbottomskip=0pt,%
          linewidth=0pt,%
          nobreak=true]%
          {keyeqn}
           
\newmdenv[font=\small,
		  linecolor=black,
          skipabove=10pt,%
          skipbelow=10pt,%
          leftmargin=0pt,%
          rightmargin=0pt,%
          innertopmargin=14pt,%
          innerbottommargin=14pt,%
          innerleftmargin=12pt,%
          innerrightmargin=12pt,%
          splittopskip=15pt,%
          splitbottomskip=5pt,%
          linewidth=0.8pt]%
          {boxedtext}

\usepackage{titlesec}          
\titleformat{\section}
{\normalfont\fontsize{15}{20}\bfseries}{\thesection}{1em}{}

\newcommand{\ob}[1]{\mkern 2mu \overline{\mkern -2mu #1 \mkern -2mu}\mkern 2mu}
\newcommand{\wt}[1]{\mkern 2mu \widetilde{\mkern -2mu #1 \mkern -2mu}\mkern 2mu}
\newcommand{\wh}[1]{\mkern 2mu \widehat{\mkern-2mu#1\mkern-2mu}\mkern 2mu}

\newcommand{\mft}[1]{\big\llbracket{#1}\big\rrbracket}

\newcommand{\fnemail}[1]{\footnote{Email: \href{mailto:#1}{\nolinkurl{#1}}}}

\begin{document}

\title{\Large\textbf{Cosmological Correlators from Resurgence\\[2mm]}}

\author{Yuanzhao Li$^{\,a\,}$\fnemail{liyuanzh22@mails.tsinghua.edu.cn}~~~~~and~~~~~Zhong-Zhi Xianyu$^{\,a,b\,}$\fnemail{zxianyu@tsinghua.edu.cn}\\[5mm]
$^a\,$\normalsize{\emph{Department of Physics, Tsinghua University, Beijing 100084, China} }\\ 
$^b\,$\normalsize{\emph{Peng Huanwu Center for Fundamental Theory, Hefei, Anhui 230026, China }} 
}

\date{}
\maketitle

\vspace{20mm}

\begin{abstract}
\vspace{10mm}

Integrating out heavy particles is generally viewed as an irreversible procedure that erases some physical information above the cutoff scale of the resulting effective theory. However, this does not preclude the possibility of recovering the heavy propagating degrees from the low-energy effective theory with appropriate boundary data supplied. A classic example is the recovery of Schwinger pair production from the resummed Euler-Heisenberg effective action. In this work, we perform a similar exercise in a cosmological setting. We show that it is possible to reconstruct the inflationary correlators with massive exchanges, for a discrete heavy spectrum and arbitrary tree topologies, from the low-energy effective theory together with unitarity and Bunch-Davies boundary conditions. The divergent EFT series is resummed in three ways: the boundary differential operators, spectral representation, and Borel resummation, and the exponentially suppressed cosmological collider signals are recovered via analytic continuation. Our construction provides a resurgent relation between the local EFT background and the nonanalytic heavy-particle production, and may also be viewed as an inverse problem for the dispersive bootstrap.
\end{abstract}

\newpage
\tableofcontents

\newpage
\section{Introduction} 

Despite the triumphant success of quantum field theories (QFTs) in quantitatively describing laws of fundamental particles and interactions in nature, our modern understanding is that all QFTs of phenomenological relevance should be treated as effective field theories (EFTs) with a finite range of validity \cite{Weinberg:1996kw}. Typically, an EFT is endowed with a power counting scheme in terms of one or more small parameters, and interaction operators are organized accordingly as a systematic expansion. It is due to this systematic expansion in small parameters that EFTs are able to make quantitative predictions with controllable precision. 

On the other hand, it has long been appreciated that the EFT expansion of physical observables is often asymptotic. If we try to compute a physical observable as a series expansion in a small parameter of EFT, the coefficients of this series often exhibit factorial growth \cite{Dyson:1952tj}. The perturbative calculations we are so used to in the ordinary Feynman-diagram treatment are thus nothing but some leading terms in an ultimately divergent series. This does not necessarily pose a practical problem, since truncating an asymptotic series is often a highly accurate numerical strategy. However, the factorial growth of the series hints at exponentially suppressed nonperturbative effects invisible at any finite order in perturbation theory. Perturbation theory, however, is not completely ignorant about them. At face value, a perturbative expansion would perceive the existence of the small nonperturbative effect through the failure of convergence at high orders. 

Asymptotic series are monsters, but they are not entirely untamable. In many cases, perturbative theory contains enough information to recover exponentially suppressed nonperturbative effects. A classic example is the Schwinger production of electron-positron pairs in a strong electric field. As Schwinger showed in \cite{Schwinger:1951nm}, the pair production can be viewed as a vacuum decay process from the viewpoint of an effective theory where the heavy particles are integrated out, and the production rate is associated with the imaginary part of the effective action. It was subsequently understood that the EFT as a weak field expansion and/or derivative expansion leads to a divergent power series for the effective action, but a proper resummation of the divergence can correctly reproduce the exponentially small effect of particle creation \cite{Dunne:1999uy,Dunne:1999vd,Dunne:2022esi}. The resummation can even be performed approximately with a finite number of Wilsonian coefficients \cite{Florio:2019hzn}. Also, the pair production can be naturally understood as an instanton solution \cite{Kim:2000un,Dunne:2005sx}. More generally, it is an interesting question that to what extent we can infer the information about physics beyond the cutoff scale from a low-energy EFT \cite{Adams:2006sv,Arkani-Hamed:2020blm,Calisto:2026pvv}.

In this work, we investigate a related question in a cosmological setting through a simple setup. Particle production during cosmic inflation is, on the one hand, very relevant to the observed large-scale inhomogeneities and anisotropies in our universe, and, on the other hand, can be thought of as a cosmic Schwinger effect triggered by a nontrivial gravitational background. Therefore, the problem of generating large-scale fluctuations has a certain nonperturbative nature in it and has been studied from many different perspectives. A standard treatment is to study the time evolution of quantum modes of heavy particles and to read off the production rate through the Bogoliubov coefficients linking the creation/annihilation operators associated with the initial vacuum state and with a late-time observer \cite{Birrell:1982ix}. The particle production during the time evolution of quantum modes can also be understood as a Stokes phenomenon, strengthening its relation to usual asymptotic analysis \cite{Sou:2021juh}. The problem was also studied from a direct EFT viewpoint where the exponentially small effect from the heavy propagating degrees is partially captured by a resummation procedure \cite{DuasoPueyo:2025lmq}. 

\begin{figure}[t]
\centering
\includegraphics[width=0.85\textwidth]{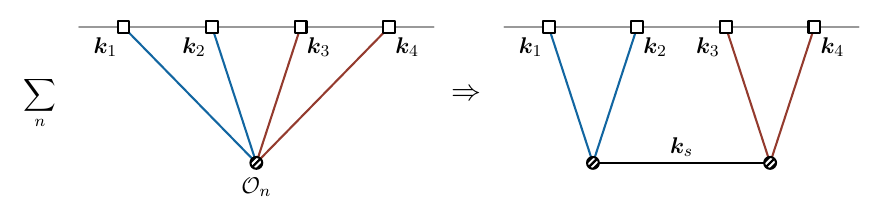}
\caption{An illustration that a proper resummation of EFT contact graphs recovers the correlator with single heavy exchange.}
\label{fig_resurgence}
\end{figure}

\begin{figure}[t]
\centering
\includegraphics[width=0.95\textwidth]{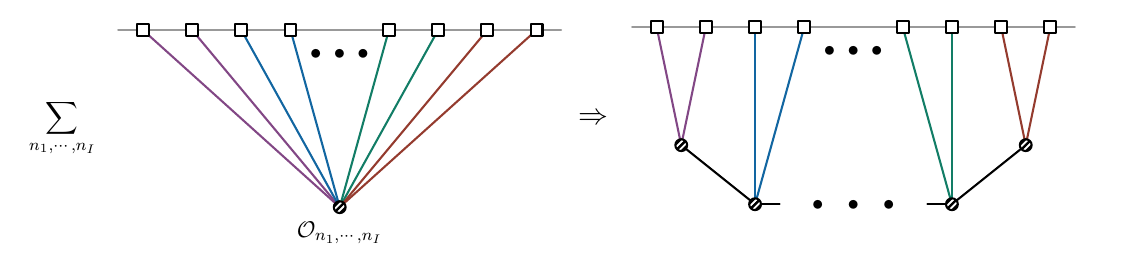}
\caption{An illustration of the resurgence of a general graph with arbitrary massive exchange and arbitrary tree topology from the EFT resummation.}
\label{fig_resurgence_general}
\end{figure}

Our approach in this work is to recover, through the lens of cosmological correlators, the exponentially small propagating effects of a heavy state from a low-energy EFT where the heavy state is integrated out.\footnote{The term of low-energy EFT should be used with caution when applied to cosmological correlators: By design, all kinematic variables specifying a correlator, i.e., the external momenta, are comoving quantities, meaning that they are essentially angles that we measure in the sky and thus carry no intrinsic scales. In this work, the low-energy EFT really means the large mass limit compared to the inflationary Hubble scale.} The setup is similar to \cite{DuasoPueyo:2025lmq}, and the new results here are the following. First, we are able to perform this exercise for an ordinary Lorentz-covariant heavy scalar particle rather than considering particles of vanishing or infinite sound speed. Second, we are able to reconstruct the full correlator in the UV, including the original EFT pieces and all types of oscillatory features coming from the propagating heavy degree, by imposing appropriate boundary conditions. Third, we perform the resummation in three complementary approaches: the boundary resummation through differential equations, the spectral resummation, and the Borel resummation. Fourth, we achieve the EFT resummation for more general cases, including a UV mass spectrum with finitely many heavy states and UV processes with arbitrary tree topologies. Finally, we further show that the  resummation can be performed approximately from truncated EFT series with Padé approximation. A graphic summary of our analysis for the simple case of single-massive exchange is shown in Fig.\;\ref{fig_resurgence}. See also Fig.\;\ref{fig_resurgence_general} for the general case. 

\paragraph{Motivations}
Besides being fun EFT gymnastics, we have three major motivations for performing this exercise at the level of cosmological correlators. 

First, the cosmological correlators with heavy-state exchanges are central observables in cosmological collider (CC) physics \cite{Chen:2009we,Chen:2009zp,Baumann:2011nk,Chen:2012ge,Noumi:2012vr,Arkani-Hamed:2015bza}. So, the question asked here is not only a matter of theoretical curiosity but also of phenomenological relevance. When doing particle-physics model building for CC physics, we routinely use QFT language for particles and interactions in the bulk inflationary spacetime \cite{Chen:2015lza,Lee:2016vti,Chen:2016uwp,Chen:2016hrz,chen:2018xck,Lu:2019tjj,Kumar:2019ebj,Liu:2019fag,Wang:2019gbi,Wang:2020ioa,Bodas:2020yho,Lu:2021wxu,Cui:2021iie,Tong:2022cdz,Reece:2022soh,Chen:2022vzh,Craig:2024qgy,Bodas:2025vpb,Anbajagane:2025uro,Kumar:2025anx,Colas:2025ind,Jiang:2025mlm,Ferreira:2026tyj,Green:2026yev,Aoki:2026olh,You:2026xoq}. Now, if all phenomenological QFTs are ultimately EFTs, it would be nice to know their reach and limitations in a more precise manner. In particular, after integrating out a heavy state, what information is preserved and what is missing in the resulting local EFTs? Cosmology brings additional nuances to these questions in that the boundary conditions, such as initial and final states, are typically not trivial compared to Minkowskian QFTs. Later, we will show that it is precisely this information about the boundary conditions that must be supplemented to the local Lagrangian of EFT in order to recover the correct propagating effects in the UV.  

Our second motivation has to do with correlators themselves. Various aspects of correlators with massive exchanges have been studied extensively in recent years with many exciting results obtained \cite{Baumann:2022jpr,Arkani-Hamed:2018kmz,Baumann:2019oyu,Sleight:2019mgd,Sleight:2019hfp,Sleight:2020obc,Sleight:2021plv,Pimentel:2022fsc,Jazayeri:2022kjy,Qin:2022fbv,Qin:2023ejc,Aoki:2023wdc,Aoki:2024uyi,Liu:2024str,Xianyu:2025lbk,Qin:2025xct,Baumann:2026atn,Arundine:2026myr,Qin:2022lva,Qin:2024gtr,Qin:2023bjk,Qin:2023nhv,Xianyu:2022jwk,Zhang:2025nzd,Grafe:2026qsm,Xianyu:2023ytd,Fan:2024iek,Fan:2025scu,Liu:2024xyi,Werth:2024mjg,Belrhali:2026ktb,Belrhali:2026rkn,Belrhali:2026jqe,Belrhali:2026act,Grafe:2026avi,Liu:2026jzn,Aoki:2026vbc,Arkani-Hamed:2023bsv,Arkani-Hamed:2023kig,Pimentel:2026kqc,Baumann:2024mvm,Cespedes:2025dnq,Bhowmick:2025mxh,Jain:2025maa,Chen:2026dqp,Huenupi:2026abj,Huenupi:2026aqc,Pinol:2026xnl,Wang:2026lff,Belrhali:2026uxn,An:2017hlx,Wang:2021qez,Werth:2023pfl,Pinol:2023oux,Werth:2024aui,Qin:2026yuo,Borinsky:2026roj}. Through these explorations, it has been increasingly clear that these correlators are not just hard to compute; they are intrinsically complicated objects. Most known correlators at tree and loop levels are expressed in terms of alien hypergeometric functions of many variables, suggesting that we should work harder to understand their properties. In particular, it would be nice to obtain different representations of these functions from different viewpoints, in different regions, if only to enrich our understanding of these objects. Indeed, many different representations and approaches have been proposed recently, and a clear picture of the essential structures of these correlators is emerging. In this context, our exercise here is to provide yet another perspective on massive correlators from the low-energy viewpoint. 

The third motivation is linked to the previous one but more technical: We want to solve the inverse problem of dispersive bootstrap \cite{Liu:2024xyi,Liu:2026jzn} of massive correlators, and thereby show a resurgent structure of cosmological correlators with massive exchanges. To explain it, we need to introduce some terminology about which we will be more precise later. Roughly, a four-point correlation function $\G(k_{12},k_{34},k_s)$ of inflaton fluctuations (or conformal scalars) with single massive exchange ($k_{12}\equiv k_1+k_2$ and $k_{34}\equiv k_3+k_4$; See Fig.\;\ref{fig_resurgence}), when restricted in the (physically reachable) parameter region $k_s<k_{34}<k_{12}$, can be broken into three pieces according to their analytic properties in kinematic variables, $\G=\G_\text{NS}+\G_\text{LS}+\G_\text{BG}$. Here, the nonlocal signal $\G_\text{NS}$, being nonanalytic in the limit of soft momentum transfer $k_s\to 0$, gives rise to the oscillatory shape in the same limit. It is a hallmark of on-shell particle production of heavy particles, and, as such, is exponentially small in the large mass limit. The local signal $\G_\text{LS}$, being analytic in small $k_s$ but nonanalytic in small $k_{34}/k_{12}$, is also a propagating effect of the heavy state and is $\sim e^{-\pi m}$ (with inflationary Hubble parameter $H=1$). On the other hand, the background $\G_\text{BG}$ is analytic in both small $k_s$ and small $k_{34}/k_{12}$. The background is essentially from the EFT limit and thus is suppressed by powers of $1/m^2$ in the large mass limit. 

The idea of dispersive bootstrap in \cite{Liu:2024xyi} is to reconstruct the whole correlator $\G$ from the signal $\G_\text{NS}+\G_\text{LS}$ or from the nonlocal signal $\G_\text{NS}$ alone. Since the nonlocal signal represents the on-shell data in a correlator in the small-momentum-transfer limit, this idea can be viewed as a dS counterpart of on-shell bootstrap of flat-space scattering amplitudes \cite{Liu:2026jzn}. Technically, this is achieved by a dispersion relation which expresses $\G$ as an integral over the discontinuity of a certain branch cut of $\G_\text{NS}$. Conceptually, it shows that the nonlocal signal $\G_\text{NS}$ contains essentially all information to recover the whole correlator $\G$, up to local terms that are fixed by UV boundary conditions, much similar to that BCFW recursion almost fixes a scattering amplitude up to local terms determined by the UV boundary. Detailed analysis in \cite{Liu:2024xyi} shows that the dispersion relation is made possible by a delicate joint cancellation of branch cuts among the three pieces in $\G$, suggesting that, probably, the background $\G_\text{BG}$ also contains enough information to recover the whole correlator $\G$. Therefore, it is conceivable that an inverse problem to dispersive bootstrap should be solvable, namely, to construct $\G$ from its background $\G_\text{BG}$ alone, supplemented by possible boundary conditions. This is exactly the EFT resummation that we study in this paper.  

\paragraph{Main results of this work} In most part of this work (Secs.\;\ref{sec_review}-\ref{sec_Borel}), we focus on the four-point correlation function of two conformal scalars of distinct flavors, with an intermediate single massive exchange. To set the stage, basic facts about the four-point correlator with single massive exchange are reviewed in Sec.\;\ref{sec_review}. 

Then, in Sec.\;\ref{sec_EFT}, we study the low-energy EFT of this model, and show that the same correlation function is generated by an infinite number of contact graphs with operators $\mathcal{O}_n$ of increasing mass dimensions. Formally, the resulting graph $\G_n$ with a single $\mathcal{O}_n$ contact insertion provides the term of $\order{m^{-2n-2}}$ of the original correlator in the large $m$ limit and we compute these contact graphs directly from their bulk time integrals in Sec.\;\ref{sec_compute_contact}. 

In Sec.\;\ref{sec_boundary}, we switch the viewpoint and review the EFT contact graphs $\G_n$ from the future boundary. Here, we show that all $\G_n$ with $n=1,2,\cdots$ can be simply constructed from the simplest contact graph $\G_0$ with direct coupling by acting $n$-times with a differential operator $\mathcal{W}_{r}$, and this differential operator is identical to the differential operator appearing in the equation for the single-massive-exchange correlator but with the mass parameter set to zero. From this viewpoint, it is very natural to see the EFT resummation as a resummation of differential operators. 

In Sec.\;\ref{sec_signal}, we show that a naive sum over all contact graphs $\G_n$ is badly divergent, but the divergence is resummable. Instead of a standard Borel resummation, we find it intuitive to adopt a spectral resummation, as detailed in Sec.\;\ref{sec_spect_resum}. In the spectral space \cite{Melville:2024ove,Werth:2024mjg}, the resummation manifests itself as a simple geometric series $\sum\limits_{n=0}^\infty z^{2n}=1/(1-z^2)$, and the new poles that emerge from this resummation (namely $z=\pm1$) give rise to the exponentially suppressed effects of particle production. Crucially, we will show in Sec.\;\ref{sec_bdry_cond} that this resummation is incomplete in that important boundary conditions of the UV theory are absent in the low-energy EFT. However, by imposing the condition of unitarity and Bunch-Davies (BD) initial state, we will recover the correct four-point correlator with single massive exchange in the UV model. 

In Sec.\;\ref{sec_Borel} we revisit the resurgence of the single-exchange correlator with the standard Borel resummation. We take the Borel transform of the EFT series for the correlator and study the analytic properties of the Borel integrand. The Borel integrand is naturally separated into two terms, each of which develops a series of poles and branch cuts. As is well known, these singularities control the nonperturbative contributions to the correlator that are hidden in the EFT series. We show that a direct Borel integral along the positive real axis leads to   the background of the UV correlator, while the signals are recovered by properly deforming the Borel contour and picking up residues/discontinuities of corresponding poles/cuts.  

In Sec.\;\ref{sec_general}, we go beyond the single-massive-exchange process and consider more general situations. We consider two directions for generalization. First, in Sec.\;\ref{sec_superPos}, we still consider the 4-point correlator with a single internal line but now allow for the exchange of $M\geq 1$ heavy states. While the exact resummation in this case is $M$ copies of the single massive exchange, this is a natural situation for us to consider approximate resummation from a truncated EFT series. With the spectral representation, we show that the standard Padé approximation can be imported directly from the more well-known flat-space realization. Then, in Sec.\;\ref{sec_generalTree}, we consider the most general tree topologies, and show that the EFT resummation still works in this case, leading to the known massive-family-tree representation for the analytic part of the full result. Then, as known in the literature, the appropriate boundary conditions fix the whole correlator through the cuts of these massive family trees (MFTs). This is schematically shown in Fig.\;\ref{fig_resurgence_general}.

We collect further discussions and outlooks  in Sec.\;\ref{sec_conclusion} and a few mathematical definitions and detailed derivations in the three appendices.

\paragraph{Notations and conventions} We work with the inflation patch of the dS spacetime with coordinates $x^\mu=(\tau,\bm x)$ and metric $\di s^2=(-\di\tau^2+\di\bm x^2)/\tau^2$, where $\tau\in(-\infty,0)$ and $\bm x\in\mathbb{R}^3$. We fix the energy unit by setting the Hubble parameter $H=1$. We will frequently use the Euler operator $\vartheta_z\equiv z\pd_z$ for a variable $z$. We also use the condensed subscripts to denote summations, such as $k_{12}\equiv k_1+k_2$, $p_{12}=p_1+p_2$, etc. Our conventions for Euler $\Gamma$ products, the Pochhammer symbol, and hypergeometric functions are collected in App.\;\ref{app_2F1}.

\section{Review of Single-Massive-Exchange Correlator}
\label{sec_review}

The answer to the four-point correlator of conformal scalars with single massive exchange has long been known. It is useful to review this answer before we start the EFT analysis, so that readers may get a better sense of direction in the subsequent analysis. We provide a short review in this section and use the opportunity to introduce our notations and terminologies. 

For later simplicity, we will consider the correlation function of two conformal scalars $\phi_1$ and $\phi_2$, both of mass $m=\sqrt 2$ but of distinct flavors. We assume that they couple to a heavy scalar particle $\si$ of mass $m>3/2$ through direct cubic interactions with couplings of arbitrary power-law time dependences. The reason to assume $m>3/2$ is that the two scaling dimensions $\Delta_\pm$ of $\si$ at late times become complex in this case; $\Delta_\pm=\fr32\pm\ii\wt\nu$ where $\wt\nu\equiv \sqrt{m^2-9/4}$ is also called the \emph{mass parameter}. As a result, we get oscillatory signals for CC applications. Also, this is consistent with taking the large-mass limit when we perform the EFT analysis. 

The Lagrangian of our model is:
\bge
\label{eq_Lag}
  \ld =-\FR{1}{2}\Big[a^2(\pd_\mu\si)^2+a^4m^2\si^2\Big]-\FR{1}{2}\sum_{I=1}^2\Big[a^2 (\pd_\mu\phi_I)^2+2a^4 \phi_I^2+\lam_I a^{-p_I+2}\phi_I^2\si\Big],
\ede
where we have spelled out explicitly the dependence on the scale factor $a(\tau)=-1/\tau$ and therefore all Lorentz indices are raised/lowered by Minkowski metric $\eta_{\mu\nu}=\,\text{diag}\,(-,+,+,+)$. A scale invariant theory corresponds to choosing $p_1=p_2=-2$, but we leave $p_{1,2}$ unspecified to allow for more general situations. 

Then, we consider the connected four-point correlation function of conformal scalars of mixed type, $\la\phi_1(\bm k_1)\phi_1(\bm k_2)\phi_2(\bm k_3)\phi_2(\bm k_4)\ra'$, where the prime on $\la\cdots\ra'$ means that the momentum conserving $\de$-function factor is removed. At the tree level, this correlator is contributed by a single $\si$-exchange process in the $s$-channel, as shown on the right panel of Fig.\;\ref{fig_resurgence}. In subsequent sections, we will also call it the UV correlator, since it is computed from the UV theory of the low-energy EFT to be introduced below. Clearly, we consider the two-flavor model for external conformal scalars to exclude $t$- and $u$-channel contributions, which is purely a technical simplification rather than a conceptual necessity. 

The four-point correlator $\la\phi_1\phi_1\phi_2\phi_2\ra'$ has been extensively studied from many angles. Here we list several basic results about it. First, it is convenient to rewrite the four-point correlator as a dimensionless function of dimensionless variables. For this purpose, we first define $k_{12}\equiv |\bm k_1|+|\bm k_2|$, $k_{34}\equiv |\bm k_3|+|\bm k_4|$, and $k_s\equiv |\bm k_1+\bm k_2|$, and then introduce the following two independent dimensionless ratios in accordance with the literature:
\begin{align}
  &r_1\equiv \FR{k_s}{k_{12}},
  &&r_2\equiv \FR{k_s}{k_{34}},
\end{align}
In this way, we can rewrite the four-point correlator as:
\begin{align}
\label{eq_dimlessG}
  \la\phi_1\phi_1\phi_2\phi_2\ra'=\FR{\tau_f^4}{16k_1k_2k_3k_4}\FR{1}{k_{12}^{1+p_1}k_{34}^{1+p_2}k_s^3}\G(r_1,r_2),
\end{align}
where $\tau_f$ is a late-time cutoff which traces the leading late-time falloff of a conformal scalar. Here, $\G$ is a dimensionless correlator that depends on kinematic variables only through the two ratios $r_{1,2}$. From now on, we will only consider this dimensionless correlator. 

The correlator $\G$ can be computed from the Lagrangian (\ref{eq_Lag}) through the standard diagrammatic expansion in the Schwinger-Keldysh formalism. For the single $\si$-exchange graph in the right panel of Fig.\;\ref{fig_resurgence}, it boils down to the following time integral:
\begin{align}
\label{eq_GTimeInt}
  \G=\sum_{\aa_1,\aa_2=\pm}\int_{-\infty}^0\prod_{I=1}^2\Big[\di z_I\,(\ii \aa_I)(-z_I)^{p_I}e^{\ii \aa_I z_I}\Big]\wt D_{\aa_1\aa_2}(r_1z_1,r_2z_2),
\end{align}
where we have introduced the dimensionless time variables $z_1=k_{12}\tau_1$ and $z_2=k_{34}\tau_2$, and the dimensionless bulk propagators $\wt D_{\aa_1\aa_2}$ for the massive scalar $\si$ are given by:
\begin{align}
  \wt D_{-+}(z_1,z_2)=&~\FR{\pi}4e^{-\pi\wt\nu}(z_1z_2)^{3/2}\text{H}_{\ii\wt\nu}^{(1)}(-z_1)\text{H}_{-\ii\wt\nu}^{(2)}(-z_2),\\
  \wt D_{+-}(z_1,z_2)=&~\FR{\pi}4e^{-\pi\wt\nu}(z_1z_2)^{3/2}\text{H}_{-\ii\wt\nu}^{(2)}(-z_1)\text{H}_{\ii\wt\nu}^{(1)}(-z_2),\\
  \wt D_{\pm\pm}(z_1,z_2)=&~\wt D_{\mp\pm}(z_1,z_2)\theta(z_1-z_2)+\wt D_{\pm\mp}(z_1,z_2)\theta(z_2-z_1),
\end{align}
and $\text{H}_\nu^{(j)}$ is Hankel function of the $j$-th kind. Also, we suppress the trivial dependence on the coupling constants $\lam_{1,2}$ here and below when writing expressions for correlators.

\paragraph{Signals and background} The explicit result for the correlator $\G$ can be obtained in a variety of ways and is often written as hypergeometric functions. Due to the existence of a large number of transformation-of-variable identities for these functions, we can represent the result in different ways, and different representations are advantageous in different regions. For our purpose, it suffices to work in a subregion of physically reachable kinematics with $0<r_1<r_2<1$. In this region, the correlator $\G$ has the following expression:\footnote{The result for region with $r_1>r_2$ can be obtained by flipping the subscripts $1\leftrightarrow 2$ for all relevant quantities. In some previous works, the local signal and background defined for $r_1<r_2$ were denoted by $\G_{\text{LS},>}$ and $\G_{\text{BG},>}$. In this work, we drop the subscript ``$>$'' since we exclusively work with $r_1<r_2$.}
\bge
\label{eq_Gresult}
  \G(r_1,r_2)=\G_\text{NS}(r_1,r_2)+\G_{\text{LS}}(r_1,r_2)+\G_{\text{BG}}(r_1,r_2).
\ede
Here, the three terms are respectively called the nonlocal signal (NS), the local signal (LS), and the background (BG). 

The explicit expressions for the three pieces are:
\begin{align}
   \label{eq_GNS}
   \G_\text{NS}(r_1,r_2)=&~\mft{1^\sharp}\mft{2^\sharp}+\text{c.c.};\\
   \label{eq_GLS}
   \G_{\text{LS}}(r_1,r_2)=&~\mft{1^\sharp}\mft{2^\flat}+\text{c.c.};\\
   \label{eq_GBG}
   \G_{\text{BG}}(r_1,r_2)=&~\sum_{\ell,n=0}^{\infty} 
      \FR{(-1)^{\ell}4\cos(\pi p_{12}/2)\Gamma(p_{12}+\ell+2n+5)}{\ell!\big(\fr{p_2+\ell}{2}+\fr{5}{4}+\fr{\ii\wt\nu}{2}\big)_{n+1}\big(\fr{p_2+\ell}{2}+\fr{5}{4}-\fr{\ii\wt\nu}{2}\big)_{n+1}}\Big(\FR{r_1}2\Big)^{2n+3}\Big(\FR{r_1}{r_2}\Big)^{p_2+\ell+1}.
\end{align}
Here, for simplicity, we have expressed the nonlocal and local signals in terms of (tuned) MFT functions. Here $\mft{i^\sharp}$ and $\mft{i^\flat}$ $(i=1,2)$ are respectively called the augmented and flattened single-site MFT and are defined as:
\begin{align} 
\label{eq_i_sharp}
  \mft{i^\sharp}=&- \FR{2^{3/2+p_i}\cos\big[\fr{\pi(p_i+\ii\wt\nu+3/2)}{2}\big]}{\sin(\pi\ii\wt\nu)} {}_{2}\mathcal{F}_1\left[\bgm\fr{p_i}{2}+\fr{5}{4}+\fr{\ii\wt\nu}{2},\fr{p_i}{2}+\fr{7}{4}+\fr{\ii\wt\nu}{2}\\1+\ii\wt\nu\edm \middle| r_i^2 \right]r_i^{3/2+\ii\wt\nu}, \\
\label{eq_2_flat}
  \mft{i^\flat}=&~\FR{2^{3/2+p_i}\cos\big[\fr{\pi(p_i+\ii\wt\nu+3/2)}{2}\big]}{\sin(\pi\ii\wt\nu)} {}_{2}\mathcal{F}_1\left[\bgm\fr{p_i}{2}+\fr{5}{4}-\fr{\ii\wt\nu}{2},\fr{p_i}{2}+\fr{7}{4}-\fr{\ii\wt\nu}{2}\\1-\ii\wt\nu\edm \middle| r_i^2 \right]r_i^{3/2-\ii\wt\nu}.
\end{align}
As mentioned, they are expressed as univariate hypergeometric functions (together with power functions). (Here we are using the dressed version of the hypergeometric function; See App.\,\ref{app_2F1} for the definition.) In the limit $0<r_i\ll 1$, all these ${}_2\mathcal{F}_1$ functions approach constants, so that the analytic structure in these limits is fully controlled by the power factors $r_i^{3/2\pm\ii\wt\nu}$. It is clear that the nonlocal signal is nonanalytic when $r_1r_2\to 0$ (namely, when $k_s\to 0$) but the local signal is analytic in this limit. However, the local signal is nonanalytic when $r_1/r_2\to 0$ (namely $k_{34}/k_{12}\to 0$). Phenomenologically, the nonlocal and local signals generate oscillatory shapes in the primordial four-point function in $r_1r_2$ and $r_1/r_2$, respectively. For heavy states with $\wt\nu\gg 1$, it is clear that both nonlocal and local signals are suppressed by the Boltzmann factor $e^{-\pi\wt\nu}$. As a result, if we consider an EFT with the heavy particle $\si$ integrated out, we are not going to see these signals at any finite order in perturbation theory, since perturbation theory only captures power-law suppressed contributions $1/m^{2n+2}$ with $n\geq 0$ when $m \gg 1$.

 On the other hand, the background $\G_{\text{BG}}$ can be expressed as a bivariate hypergeometric function (a Kampé de Fériet function) which we do not explicitly spell out here. Instead, we show its series definition in the region of interest.\footnote{In the context of MFTs, this part is also written as $\mft{12}$.} From this explicit sum, it is clear that the background is analytic when both $r_1\to 0$ and $r_1/r_2\to 0$. More interestingly, when the massive particle is heavy, $\wt\nu\simeq m\gg 1$, the term of $\order{r_1^{2n+3}}$ scales like $m^{-2(n+1)}$. These terms are in principle what we could get from the low-energy EFT since they are power-law suppressed, and the series (\ref{eq_GBG}) looks suspiciously like an EFT expansion. However, it is not; The background series (\ref{eq_GBG}) is really an expansion in soft kinematic variables rather than in inverse powers of $\wt\nu^2$. As such, the series (\ref{eq_GBG}) is a well-defined Taylor series with a finite convergence domain. On the contrary, as we shall show below, the EFT expansion of the correlator in terms of $1/m^2$ is actually divergent. This divergence should be expected, since the Boltzmann factors from nonlocal and local signals show that the $m\to\infty$ limit is an intrinsic singularity of the correlator.

\paragraph{Differential Equations} The dimensionless correlator $\G$ satisfies a set of differential equations:
\begin{align}
\label{eq_DE1ofG}
  \mathcal{D}_{r_1}^{(\wt\nu, p_1)}\G=&~\FR{2r_1^{p_2+4}r_2^{p_1+4}}{r_{12}^{p_{12}+5}}\cos(\pi p_{12}/2)\Gamma(p_{12}+5),\\
\label{eq_DE2ofG}
  \mathcal{D}_{r_2}^{(\wt\nu, p_2)}\G=&~\FR{2r_1^{p_2+4}r_2^{p_1+4}}{r_{12}^{p_{12}+5}}\cos(\pi p_{12}/2)\Gamma(p_{12}+5),
\end{align}
where the differential operator $\mathcal{D}_{r}^{(\wt\nu, p)}$ is defined by:
\bge
\label{eq_Dr}
  \mathcal{D}_{r}^{(\wt\nu,p)}\equiv \Big(\vartheta_{r}-\FR32\Big)^2+\wt\nu^2-r^2(\vartheta_{r}+p+2)(\vartheta_{r}+p+1).
\ede
Here and below, we use the Euler operator $\vartheta_z\equiv z\pd_z$. Also, we use the shorthands $p_{12}=p_1+p_2$ and $r_{12}=r_1+r_2$.   
From the viewpoint of differential equations, the background $\G_{\text{BG}}$ by itself solves the above two inhomogeneous equations with the additional condition that it is analytic when both $r_1\to 0$ and $r_1/r_2\to 0$. On the other hand, to get the complete correlator $\G$, the background $\G_{\text{BG}}$ should be supplemented with homogeneous solutions which are fixed by boundary conditions. After imposing the usual BD initial condition for both the massive scalar $\si$ and the conformal scalars $\phi_{1,2}$, the homogeneous solutions are uniquely fixed to be the sum of nonlocal and local signals.

\paragraph{Dispersive Bootstrap} From the explicit expressions (\ref{eq_GNS})-(\ref{eq_GBG}) for the correlator $\G$, it is evident that there is a sense of simplicity for $\G_\text{NS}$ and $\G_\text{LS}$ compared to $\G_\text{BG}$. The underlying physics is that the signals are generated by on-shell produced particles. As a result, signals are associated with cutting rules and factorization theorems. This is particularly well understood for the nonlocal signal, which comes from the region where the heavy particle $\si$ is produced pairwise and stretched to superhorizon distances. So, it represents the on-shell part of the correlator when the momentum transfer $k_s$ goes to zero. This motivates an on-shell approach for bootstrapping correlators. Due to the fact that the nonanalytic on-shell behavior of a correlator is typically associated with a branch cut, this on-shell bootstrap should be realized technically by a dispersion relation. This is the motivation for the dispersive bootstrap developed in \cite{Liu:2024xyi,Liu:2026jzn}.

At the heart of the dispersive bootstrap is an analysis of the singularity structure of $\G$. For the current presentation, it is useful to present the result by treating the correlator $\G$ as a function of $r_1$ and $x\equiv r_1/r_2$. We will write it as $\G(r_1;x)\equiv\G(r_1,r_1/x)$ and it should not be confused with the previously defined $\G(r_1,r_2)$. For the parameter space of our interest, $r_1<r_2$, we have $0<x<1$. For fixed $x$ in this region, it has been analyzed in detail that $\G(r_1;x)$ as a function of complex $r_1$ is analytic everywhere on the complex $r_1$-plane except for a branch cut on the entire negative real axis. Somewhat remarkably, the discontinuity of $\G$ across this branch cut can be entirely represented in terms of the on-shell part alone, namely the nonlocal signal $\G_\text{NS}$:
\begin{align}
\label{eq_DiscG}
  \mathop{\text{Disc}}_{r_1}\G(r_1;x)=\mathop{\text{Disc}}_{r_1}\Big[\G_\text{NS}(r_1;x)+\G_\text{NS}(-r_1;x)\Big]\theta(-r_1).
\end{align}
It means that the nonlocal signal alone contains enough information to recover the whole correlator $\G$, up to completely analytic local contributions, which are fixed by a boundary condition in the UV, namely, at $|r_1|\to\infty$. 

Something remarkable happens if we break the correlator $\G$ into the sum $\G_\text{NS}+\G_\text{LS}+\G_\text{BG}$, study the singularity of each piece, and see how they combine into the total discontinuity in $\G$ as shown in (\ref{eq_DiscG}). In short, each of the three pieces has a branch cut on the positive real $r_1$-axis, but they all combine to zero in a very ``clever'' way. While this is a natural consequence of choosing the BD initial condition and hence the cancellation of spurious folded poles, it does show that the three pieces ``know'' each other through this cancellation. It is then conceivable that we can start from one piece and reconstruct the whole correlator with certain assumptions about the boundary condition and analyticity. The dispersive bootstrap is then a program of reconstructing the correlator from the signal alone, and the EFT resummation in this work aims at reconstructing the correlator from the background alone. From this viewpoint, the EFT resummation is the inverse problem of the dispersive approach.

\section{EFT Correlator}
\label{sec_EFT}

In this section, we consider the low-energy EFT of our previous model after integrating out the heavy scalar $\si$.\footnote{To remove any confusion, we are not considering a Wilsonian EFT where all modes above an energy cutoff $\Lambda$ are integrated out. Rather, we only integrate out all modes of $\si$ and all modes of $\phi_{1,2}$ are retained.} We will obtain the low-energy effective Lagrangian and use it to recompute the correlator $\G$ introduced in the last section. Clearly, the low-energy EFT contains an infinite number of interaction operators and each of them will make a contribution to $\G$ at the tree level through a contact graph. We will work out all these contact graphs in this section. In the end, it will be clear that the series from summing over all these graphs diverges for any finite value of $1/m$. 

Let us first derive the EFT of our model by integrating out $\si$. Since the original Lagrangian (\ref{eq_Lag}) is at most quadratic in $\si$, the integrating-out procedure can be done in closed form. Conveniently, we can simply solve the classical equation of motion for $\si$, which in the inflationary coordinates $(\tau,x^i)$ becomes:
\bge
\label{eq_EoMsigma}
  \bigg[\Big(\vartheta_\tau-\FR32\Big)^2-\tau^2\pd_i^2+\wt\nu^2\bigg]\si=-\FR12\sum_{I=1}^2\lam_I (-\tau)^{p_I+2}\phi_I^2,
\ede
where $\wt\nu=\sqrt{m^2-9/4}$. Suppose that we are working in the region where all relevant physical energies (or momenta) are much less than $m$. In this case, we can take the inverse of Klein-Gordon operator and expand it in the large $\wt\nu$ limit, so that the solution to the above equation is:
\bge
  \si=-\FR12 \sum_{n=0}^\infty\FR{(-1)^n}{\wt\nu^{2n+2}}\bigg[\Big(\vartheta_\tau-\FR32\Big)^2-\tau^2\pd_i^2\bigg]^n\sum_{I=1}^2\lam_I (-\tau)^{p_I+2}\phi_I^2.
\ede
Putting it back to the original Lagrangian, we get the low-energy effective Lagrangian:
\begin{align}
\label{eq_LagEFT}
  \ld_\text{EFT}
  =-\FR{1}{2}\sum_{I=1}^2\Big[a^2 (\pd_\mu\phi_I)^2+2a^4 \phi_I^2 \Big]+\sum_{n=0}^\infty\Big[\mathcal{O}_n+\mathcal{O}^{(1)}_n+\mathcal{O}^{(2)}_n\Big],
\end{align}
where $\mathcal{O}_n$ is an infinite tower of local interaction operators that couple $\phi_1$ to $\phi_2$:
\begin{align}
   \mathcal{O}_n=\FR{\lam_1\lam_2}{4}a^4 \FR{(-1)^n}{\wt\nu^{2n+2}}(-\tau)^{p_1+2}\phi_1^2\bigg[\Big(\vartheta_\tau-\FR32\Big)^2-\tau^2\pd_i^2\bigg]^n\Big[(-\tau)^{p_2+2}\phi_2^2\Big],
\end{align}
while $\mathcal{O}_n^{(1)}$ and $\mathcal{O}_n^{(2)}$ are quartic self-couplings of $\phi_1$ and $\phi_2$, respectively:
\begin{align}
   \mathcal{O}^{(1)}_n=\FR{\lam_1^2}{8}a^4 \FR{(-1)^n}{\wt\nu^{2n+2}}(-\tau)^{p_1+2}\phi_1^2\bigg[\Big(\vartheta_\tau-\FR32\Big)^2-\tau^2\pd_i^2\bigg]^n\Big[(-\tau)^{p_1+2}\phi_1^2\Big];\\
   \mathcal{O}^{(2)}_n=\FR{\lam_2^2}{8}a^4 \FR{(-1)^n}{\wt\nu^{2n+2}}(-\tau)^{p_2+2}\phi_2^2\bigg[\Big(\vartheta_\tau-\FR32\Big)^2-\tau^2\pd_i^2\bigg]^n\Big[(-\tau)^{p_2+2}\phi_2^2\Big].
\end{align}
The self-coupling operators $\mathcal{O}_n^{(1)}$ and $\mathcal{O}_n^{(2)}$ do not contribute to $\la\phi_1\phi_1\phi_2\phi_2\ra'$ at the tree level, so we will neglect them from now on, and focus only on $\mathcal{O}_n$. 

Next, we consider the four-point correlator $\G$ introduced in (\ref{eq_dimlessG}) from the viewpoint of low-energy EFT with the Lagrangian in (\ref{eq_LagEFT}).\footnote{Obviously, from the viewpoint of the EFT, the normalization for defining $\G$ in (\ref{eq_dimlessG}) is weird, but we will stick to this normalization since a normalization is only a normalization. We prefer not to confuse readers by using multiple normalization conventions for the same quantity.} Here, at the tree level, every operator $\mathcal{O}_n$ makes a contact-graph contribution to $\G$ which we denote as $\G_n$, as shown in the left panel of Fig.\;\ref{fig_resurgence}. By the usual diagrammatic rule, it is straightforward to write down a time-integral representation for $\G_n$:
\begin{align}
\label{eq_GnTimeInt}
  \G_n(r_1,r_2)=\sum_{\aa=\pm}\ii\aa \FR{(-1)^n}{\wt\nu^{2n+2}}r_2^3\int_{-\infty}^0\di z\,(-z)^{p_1}e^{\ii\aa z}\,\mathscr{D}_{r_1z}^n\Big[(-r_1z/r_2)^{p_2+4}e^{\ii\aa r_1z/r_2}\Big].
\end{align}
Here, we have made all quantities dimensionless, with two momentum ratios defined as before and the dimensionless time $z\equiv k_{12}\tau$, where $\tau$ is the time variable for the vertex. Also, for simplicity, we have defined a differential operator:
\bge
\label{eq_Dz}
  \mathscr{D}_{r_1z}\equiv \Big(\vartheta_z-\FR32\Big)^2+r_1^2z^2.
\ede
In principle, this is a single-layer time integral over a sum of products of a plane-wave factor and a power function and thus can be carried out directly. However, the actual calculation can be quite cumbersome due to the $n$-fold action of the differential operator $\mathscr{D}_{r_1z}$. Here, let us show the result first and then explain how to derive it:
\begin{keyeqn}
\begin{align}
\label{eq_Gn}
    \G_n=&\sum_{\ell=0}^{\infty}\left(\frac{\ell+p_2+5/2}{\wt\nu}\right)^{2n}\frac{2(-1)^{n+\ell}\cos\left(\pi p_{12}/2\right)\Gamma(\ell+p_{12}+5)}{\wt\nu^2\ell!}r_1^3\left(\frac{r_1}{r_2}\right)^{\ell+p_2+1}\n\\
    &\times{}_2\mathrm{F}_1\left[\bgm \frac{\ell+p_{12}+5}{2},\frac{\ell+p_{12}+6}{2}\\\ell+p_2+\frac{7}{2}\edm\Bigg|r_1^2\right]{}_2\mathrm{F}_1\left[\bgm -\frac{\ell}{2},-\frac{\ell-1}{2}\\-\ell-p_2-\frac{3}{2}\edm\Bigg|r_2^2\right].
\end{align}
\end{keyeqn}
It is interesting to compare this result to the $n$-th order term of $\G_{\text{BG}}$ in (\ref{eq_GBG}). Both expressions are of order $1/\wt\nu^{2n+2}$. However, the difference is that $\G_n$ is exactly proportional to $1/\wt\nu^{2n+2}$, while the $n$-th order term of $\G_{\text{BG}}$ has more complicated dependence on $\wt\nu$ which goes like $1/\wt\nu^{2n+2}$ only in the large $\wt\nu$ limit. Therefore, $\G_n$ is the $n$-th order term of a genuine EFT expansion in terms of $1/\wt\nu^2$ but the $n$-th order term in $\G_{\text{BG}}$ is not. Below, we will also recast the background $\G_{\text{BG}}$ into a form that can be compared more readily to $\G_n$. However, before doing so, let us first discuss the computation of $\G_n$.

\subsection{Computing the contact graph}
\label{sec_compute_contact}
There is a nice way of doing the time integral (\ref{eq_GnTimeInt}) by using the eigenfunctions of $\mathscr{D}_z$. This approach may look like a detour, but this detour will show its value for later EFT resummation. So, let us introduce it here. 

\paragraph{Spectral representation}
Let $S(\mu;z)$ be an eigenfunction of $\mathscr{D}_z$   with eigenvalue $-\mu^2$. That is, $S(\mu;z)$ solves the following equation:
\bge
  \mathscr{D}_z S(\mu;z)=-\mu^2 S(\mu;z).
\ede
Note that the operator $\mathscr{D}_z+\mu^2$ is nothing but the Klein-Gordon operator in (\ref{eq_EoMsigma}) written in the momentum space with $\wt\nu\to \mu$. So the above equation is nothing but the classical equation of motion for a massive mode function with mass parameter $\mu$. With the usual canonical normalization, we can get a pair of independent solutions as:
\begin{align}
  &S_{+}(\mu;z)=\frac{\sqrt{\pi}}{2}e^{-\pi\mu/2}(-z)^{3/2}\mathrm{H}_{-\ii\mu}^{(2)}(-z);
  &&S_{-}(\mu;z)=S_{+}^*(\mu;z).
\end{align}
In anticipation of recovering propagating degrees, we choose to use the following orthonormal conditions for these eigenfunctions \cite{Grafe:2026qsm,Belrhali:2026rkn}:
\bge
\label{eq_OrthNormS}
  \de(x-y)=-\int_{-\infty}^{+\infty}\di \mu\,\FR{\mu\sinh(\pi\mu)}{\pi(-y)^4} S_\pm(\mu;x)S_\pm(\mu;y).
\ede
Note that this condition holds separately for $S_+$ and $S_-$. So, we rewrite the dimensionless time variable $z\to z_1$ and use  the following identity to replace the quantity in square brackets in (\ref{eq_GnTimeInt}):
\bge
(-r_1z_1/r_2)^{p_2+4}e^{\ii\aa r_1z_1/r_2}=r_2\int_{-\infty}^0\di z_2\,\de(r_1z_1-r_2z_2)(-z_2)^{p_2+4}e^{\ii\aa z_2}.
\ede
Then:
\begin{align}
    \G_n =&~\sum_{\aa=\pm} \ii\aa \FR{(-1)^{n}}{\wt\nu^{2n+2}}\int_{-\infty}^0\di z_1(-z_1)^{p_1}e^{\ii\aa z_1}\mathscr{D}_{r_1z_1}^n\bigg[\int_{-\infty}^0\di z_2\,\de(r_1z_1-r_2z_2)(-z_2)^{p_2}(-r_2z_2)^4e^{\ii\aa z_2}\bigg].
\end{align}
Then, using the $S_+$ and $S_-$-version of the identity (\ref{eq_OrthNormS}) to replace the $\de$-function in the $+$ and $-$ SK branches, respectively, we get:  
\begin{align}
    \G_n  =&~\sum_{\aa=\pm} \ii\aa \FR{(-1)^{n+1}}{\wt\nu^{2n+2}}\int_{-\infty}^{+\infty}\di \mu\, \FR{\mu\sinh(\pi\mu)}{\pi}\int_{-\infty}^{0}\di z_1\,(-z_1)^{p_1}e^{\ii\aa z_1}\mathscr{D}_{r_1z_1}^nS_{\aa}(\mu;r_1z_1)\n\\
    &\times\int_{-\infty}^{0}\di z_2\,(-z_2)^{p_2}e^{\ii\aa z_2}S_{\aa}(\mu;r_2z_2).
\end{align}
Clearly, the introduction of $S_\pm$ function trivializes the action of $\mathscr{D}_z$. Using $\mathscr{D}_{r_1z_1}^n S_\pm(\mu;r_1z_1)=(-\mu^2)^n S_\pm(\mu;r_1z_1)$, we get the following expression for $\G_n$ in which the time integral is traded for a spectral integral over $\mu$:
\begin{align}
\label{eq_RepU}
   \G_n  =&\sum_{\aa=\pm} \ii\aa \int_{-\infty}^{+\infty}\di \mu\,\rho_n(\mu)\mathcal{U}^{p_1}_\aa(\mu;r_1)\mathcal{U}^{p_2}_\aa(\mu;r_2),
\end{align}
where $\rho_n(\mu)$ is defined as follows and acts like a spectral density:
\begin{align}
\label{eq_rho_n}
    \rho_n(\mu)\equiv-\frac{\mu\sinh(\pi\mu)}{\pi\wt\nu^2}\left(\frac{\mu}{\wt\nu}\right)^{2n}.
\end{align}
On the other hand, the function $\mathcal{U}^{p}_\aa(\mu;r)$ comes from time integrals which can be finished as:
\begin{align}
\label{eq_3pt}
   & \mathcal{U}^{p}_\aa(\mu;r)\equiv\int_{-\infty}^{0}\di z(-z)^{p}e^{\ii\aa z}S_{\aa}(\mu;rz)\n\\
   & =\sqrt{\frac{2}{\pi}}e^{-\ii\aa(p+3/2)\pi/2}\sum_{\cc=\pm}\Gamma[-\ii\cc\mu,p+\ii\cc\mu+\fr{5}{2}]\left(\frac{r}{2}\right)^{\ii\cc\mu+3/2}{}_2\mathrm{F}_1\left[\bgm \frac{p+\ii\cc\mu}{2}+\frac{5}{4},\frac{p+\ii\cc\mu}{2}+\frac{7}{4}\\1+\ii\cc\mu\edm\Bigg|r^2\right].
\end{align}

\paragraph{Finishing the spectral integral} With the above spectral representation obtained, we can compute $\G_n$ by appropriately closing the integration contour of $\mu$ and collecting all poles inside, thanks to the meromorphic nature of the spectral integrand. 

To close the integration contour properly, we first finish the sum over the SK index $\aa=\pm$ as:
\begin{align}
    \G_n(r_1,r_2)=&~\FR{4\cos(\pi p_{12}/2)}{\pi^2\wt\nu^2}\int_{-\infty}^{+\infty}\di \mu\, \mu\sinh(\pi\mu)\left(\frac{\mu}{\wt\nu}\right)^{2n}\n\\
    &\times\sum_{\cc_1,\cc_2=\pm}\Gamma[-\ii\cc_1\mu,-\ii\cc_2\mu,p_1+\ii\cc_1\mu+\fr{5}{2},p_2+\ii\cc_2\mu+\fr{5}{2}]\left(\frac{r_1}{2}\right)^{\ii\cc_1\mu+3/2}\left(\frac{r_2}{2}\right)^{\ii\cc_2\mu+3/2}\n\\
    &\times{}_2\mathrm{F}_1\left[\bgm \frac{p_1+\ii\cc_1\mu}{2}+\frac{5}{4},\frac{p_1+\ii\cc_1\mu}{2}+\frac{7}{4}\\1+\ii\cc_1\mu\edm\Bigg|r_1^2\right]{}_2\mathrm{F}_1\left[\bgm \frac{p_2+\ii\cc_2\mu}{2}+\frac{5}{4},\frac{p_2+\ii\cc_2\mu}{2}+\frac{7}{4}\\1+\ii\cc_2\mu\edm\Bigg|r_2^2\right].
\end{align}
We want to close the contour in such a way that the integrand decays fast enough on the large arc. To see the behavior of the integrand at large $\mu$, the following asymptotic behavior of the hypergeometric function is useful:
\begin{align}
\label{eq_asymp}
     {}_2\mathrm{F}_1\left[\bgm \frac{p+\ii\cc\mu}{2}+\frac{5}{4},\frac{p+\ii\cc\mu}{2}+\frac{7}{4}\\1+\ii\cc\mu\edm\Bigg|r^2\right]\sim \left(\frac{r}{2}\right)^{-\ii\cc\mu}e^{-\ii\cc\mu\xi(r)}.~~~(|\mu|\to\infty)
\end{align}
where $\xi(r)\equiv\text{arccosh}\,(1/r)$ is positive for physical $r$. Therefore, the four terms with $\cc_1,\cc_2=\pm$ in the integrand have different asymptotic behaviors as $|\mu|\to \infty$, and we should close the integration contour in different ways, in either the upper or the lower half-plane. Below we denote contributions from these 4 parts as $\G_n^{(\cc_1\cc_2)}$ and compute them in turn. 

First, we consider $\cc_1=\cc_2=\cc$, where the integrand scales as $e^{-\ii\cc\mu[\xi(r_1)+\xi(r_2)]}$ as $\mu\to\infty$ up to a power function of $\mu$. So, we should close the contour from the side where $\Im(\cc\mu)<0$. In this case, there is only one set of poles coming from a $\Gamma$ factor:\footnote{The would-be $\Gamma$ pole at $\mu=0$ is canceled by the zero from the factor $\mu\sinh(\pi \mu)$.}
\begin{align}
    \mu=-\ii \cc(\ell+1).~~~(\ell=0,1,2,\cdots)
\end{align}
Collecting the residues at all these poles, we get: 
\begin{align}
    \G_n^{(\pm\pm)}=&\sum_{\ell=0}^{\infty}\frac{8(-1)^{n+\ell+1}\cos(\pi p_{12}/2)\Gamma[p_1+\ell+\fr{7}{2},p_2+\ell+\fr{7}{2}]}{\wt\nu^2(\ell+1)!\ell!}\left(\FR{\ell+1}{\wt\nu}\right)^{2n}\n\\
    &\times\left(\frac{r_1r_2}{4}\right)^{\ell+5/2}{}_2\mathrm{F}_1\left[\bgm \frac{p_1+\ell}{2}+\frac{7}{4},\frac{p_1+\ell}{2}+\frac{9}{4}\\2+\ell\edm\Bigg|r_1^2\right]{}_2\mathrm{F}_1\left[\bgm \frac{p_2+\ell}{2}+\frac{7}{4},\frac{p_2+\ell}{2}+\frac{9}{4}\\2+\ell\edm\Bigg|r_2^2\right].
\end{align}

Next, we consider the cases with $\cc_1=-\cc_2=\cc$. According to (\ref{eq_asymp}), when $|\mu|\to\infty$, the integrand behaves like $e^{-\ii\cc\mu[\xi(r_1)-\xi(r_2)]}$. Without loss of generality, we always assume $r_1<r_2$ here and below. Then, again, we should pick up all poles in the half complex plane of $\mu$ with $\Im(\cc\mu)<0$. This time there are two classes of poles:
\begin{align}
    &\mu=-\ii \cc(\ell+1),&&\mu=-\ii\cc(\ell+p_2+\fr{5}{2}).~~~(\ell=0,1,2,\cdots)
\end{align}
The first one comes from the hypergeometric function and the second from a $\Gamma$ factor.\footnote{Recall that a Gauss hypergeometric function can be written in a regularized form ${}_2\text{F}_1[\genfrac{}{}{0pt}{}{a~b}{c}|z]=\Gamma(c){}_2\wt{\text{F}}_1[\genfrac{}{}{0pt}{}{a~b}{c}|z]$ where the regularized version ${}_2\wt{\text{F}}_1$ is an entire function of $a$, $b$, and $c$.} In fact, the contributions of the first class of poles give us exactly $-\G_n^{(\pm\pm)}$. Also including the contributions from the second class of poles at $\mu=-\ii\cc(\ell+p_2+\frac{5}{2})$, we get:
\begin{align}
    \G_n^{(\pm\mp)}=-\G_n^{(\pm\pm)}+\sum_{\ell=0}^{\infty}&\left(\frac{\ell+p_2+5/2}{\wt\nu}\right)^{2n}\frac{(-1)^{n+\ell}\cos\left(\pi p_{12}/2\right)\Gamma(\ell+p_{12}+5)}{\wt\nu^2\ell!}r_1^3\left(\frac{r_1}{r_2}\right)^{\ell+p_2+1}\n\\
    &\times{}_2\mathrm{F}_1\left[\bgm \frac{\ell+p_{12}+5}{2},\frac{\ell+p_{12}+6}{2}\\\ell+p_2+\frac{7}{2}\edm\Bigg|r_1^2\right]{}_2\mathrm{F}_1\left[\bgm -\frac{\ell}{2},-\frac{\ell-1}{2}\\-\ell-p_2-\frac{3}{2}\edm\Bigg|r_2^2\right].
\end{align}
Now, we get the final answer for $\G_n$ by summing over the four contributions:
\begin{align} 
    \G_n=\sum_{\cc_1,\cc_2=\pm}\G_n^{(\cc_1\cc_2)},
\end{align}
and the result is given in (\ref{eq_Gn}). 

\section{A Boundary Perspective of EFT Resummation}
\label{sec_boundary}

In the previous section, we computed the contact-graph contribution to the four-point correlator $\G$ at each order of $1/\wt\nu^2$-expansion in the low-energy EFT. We did the computation by directly finishing the time integral with the help of a spectral trick. Before considering the resummation of these contact graphs below, it is also useful to take a boundary viewpoint for these contact processes in the EFT. In this section, we take this detour and show that all contact graphs derived above can also be obtained by acting with appropriate differential operators on the lowest-order contact graph. Interestingly, this representation suggests a way to perform EFT resummation, which is resumming the differential operators instead of resumming graphs. As we shall see, the resummation of differential operators immediately leads to the differential equation satisfied by the original UV correlator $\G$ with a heavy particle exchange.

The treatment is similar to deriving the differential equations for tree-level correlators from their SK time integrals. Our starting point is the bulk time integral for $\G_n$ in (\ref{eq_GnTimeInt}), which can be rewritten as:
\bge
  \G_n(r_1,r_2)=\sum_{\aa=\pm}\ii\aa \FR{(-1)^n}{\wt\nu^{2n+2}}r_2^3\int_{-\infty}^0\di z\,(-z)^{p_1}e^{\ii\aa z}\,f_n(r_1z),
\ede
with $f_n(r_1z)\equiv \mathscr{D}_{r_1z}^n\big[(-r_1z/r_2)^{p_2+4}e^{\ii\aa r_1z/r_2}\big]$ satisfying $f_{n}(r_1z)=\mathscr{D}_{r_1z} f_{n-1}(r_1z)$. Then, we have:
\begin{align}
  \G_n(r_1,r_2)=&\sum_{\aa=\pm}\ii\aa \FR{(-1)^n}{\wt\nu^{2n+2}}r_2^3\int_{-\infty}^0\di z\,(-z)^{p_1}e^{\ii\aa z}\mathscr{D}_{r_1z} f_{n-1}(r_1z)\n\\
  =&~\mathcal{W}_{r_1}\sum_{\aa=\pm}\ii\aa \FR{(-1)^n}{\wt\nu^{2n+2}}r_2^3\int_{-\infty}^0\di z\,(-z)^{p_1}e^{\ii\aa z} f_{n-1}(r_1z)
  =\mathcal{W}_{r_1}\left[\frac{-1}{\wt\nu^2}\G_{n-1}(r_1,r_2)\right],
\end{align}
where the differential operator $\mathcal{W}_{r_1}$ happens to be the one defined in (\ref{eq_Dr}) with $\wt\nu=0$:
\begin{align}
    \mathcal{W}_{r_1}=\mathcal{D}_{r_1}^{(0,p_1)}= \Big(\vartheta_{r_1}-\FR32\Big)^2-r_1^2(\vartheta_{r_1}+p_1+1)(\vartheta_{r_1}+p_1+2) .
\end{align}
In this way, we can use $\mathcal{W}_{r_1}$ to recursively generate all $\G_n(r_1,r_2)$ from $\G_0(r_1,r_2)$:
\begin{keyeqn}
\begin{align}
\label{eq_nDE}
    \G_n(r_1,r_2)=\FR{(-1)^n}{\wt\nu^{2n}} \mathcal{W}_{r_1}^n\G_0(r_1,r_2),
\end{align}
\end{keyeqn}
and the correlator $\G_0$ is just the familiar contact graph with a direct coupling:
\begin{align}
    \G_0(r_1,r_2)=&~\FR{1}{\wt\nu^{2}}r_1^3\left(\FR{r_1}{r_2}\right)^{p_2+1}\sum_{\aa=\pm} \int_{-\infty}^0\di z(\ii\aa)(-z)^{p_{12}+4}e^{\ii\aa (1+r_1/r_2)z}\n\\
    =&~\frac{2}{\wt\nu^2}\cos(\pi p_{12}/2)\Gamma(p_{12}+5)\frac{r_1^{p_2+4}r_2^{p_1+4}}{r_{12}^{p_{12}+5}}.
\end{align}

As we shall show more explicitly in the next section, a naive sum over all $\G_n$ is badly divergent, but the above rewriting of contact graphs $\G_n$ immediately suggests a way of resummation:
\begin{align}
\label{eq_inverseD}
    \G_\text{resum}(r_1,r_2)\sim\sum_{n=0}^{\infty}\G_{n}(r_1,r_2)=\bigg[\sum_{n=0}^\infty\FR{(-1)^n}{\wt\nu^{2n}}\mathcal{W}_{r_1}^n\bigg]\G_{0}(r_1,r_2)\sim\left(\mathcal{W}_{r_1}+\wt\nu^2\right)^{-1}\wt\nu^2\G_{0}(r_1,r_2).
\end{align}
Applying the operator $\mathcal{D}_{r_1}^{\wt\nu, p_1}=\mathcal{W}_{r_1}+\wt\nu^2$ on both sides of (\ref{eq_inverseD}), we get nothing but the original differential equation (\ref{eq_DE1ofG}) for the full correlator $\G$:
\begin{align}
    \mathcal{D}_{r_1}^{(\wt\nu,p_1)}\G_\text{resum}(r_1,r_2)=2\cos(\pi p_{12}/2)\Gamma(p_{12}+5)\frac{r_1^{p_2+4}r_2^{p_1+4}}{r_{12}^{p_{12}+5}}.
\end{align}
Clearly, a similar treatment also leads to the second differential equation (\ref{eq_DE2ofG}). In total, the result here shows that the resummed EFT graph is identical to the original correlator with heavy particle exchange up to boundary conditions that cannot be seen from the differential equations alone. Once these boundary conditions are properly included, the EFT resummation will lead to an answer that matches the UV calculation. Using the boundary condition to fix the homogeneous solution is a familiar exercise in cosmological bootstrap and thus we will not repeat the story here. We will be more explicit about these boundary conditions in the next section when reconstructing the correlator by directly resumming $\G_n$ without using differential equations. 
 
Finally, it is also possible to derive explicit expressions for all $\G_n$ from (\ref{eq_nDE}). The starting point is to recognize that the leading contact graph $\G_0$ can be written in a spectral form:
\begin{align}
\label{eq_G0}
    \G_0(r_1,r_2)
    =&~\frac{2}{\wt\nu^2}\cos(\pi p_{12}/2)\Gamma(p_{12}+5)\frac{r_1^{p_2+4}r_2^{p_1+4}}{r_{12}^{p_{12}+5}}\n\\
    =&\sum_{\ell=0}^{\infty}\frac{2(-1)^{\ell}\cos\left(\pi p_{12}/2\right)\Gamma(\ell+p_{12}+5)}{\wt\nu^2\ell!}\n\\
    &\times r_1^3\left(\frac{r_1}{r_2}\right)^{\ell+p_2+1}{}_2\mathrm{F}_1\left[\bgm \frac{\ell+p_{12}+5}{2},\frac{\ell+p_{12}+6}{2}\\\ell+p_2+\frac{7}{2}\edm\Bigg|r_1^2\right]{}_2\mathrm{F}_1\left[\bgm -\frac{\ell}{2},-\frac{\ell-1}{2}\\-\ell-p_2-\frac{3}{2}\edm\Bigg|r_2^2\right].
\end{align}
This is of course obtained by setting $n=0$ in (\ref{eq_Gn}), but it can also be proved more directly without resorting to a spectral representation. We provide a proof in App.\,\ref{app_G0}. Once again, this representation trivializes the action of differential operator $\mathcal{W}_{r_1}$, because the summand in (\ref{eq_G0}) is the eigenfunction of the differential operator $\mathcal{W}_{r_1}$: 
\begin{align}
    &\mathcal{W}_{r_1}\left\{r_1^{\ell+p_2+4}{}_2\mathrm{F}_1\left[\bgm \frac{\ell+p_{12}+5}{2},\frac{\ell+p_{12}+6}{2}\\\ell+p_2+\frac{7}{2}\edm\Bigg|r_1^2\right]\right\}\n\\
    &=(\ell+p_2+5/2)^2\left\{r_1^{\ell+p_2+4}{}_2\mathrm{F}_1\left[\bgm \frac{\ell+p_{12}+5}{2},\frac{\ell+p_{12}+6}{2}\\\ell+p_2+\frac{7}{2}\edm\Bigg|r_1^2\right]\right\},
\end{align}
Then, the $n$-fold action of $\mathcal{W}_{r_1}$ on (\ref{eq_G0}) reproduces the spectral representation of $\G_n(r_1,r_2)$ in (\ref{eq_Gn}).

\section{Resurgence with Spectral Resummation}
\label{sec_signal}
In the previous two sections, we have constructed the tree-level four-point correlator $\G_n$ contributed by every contact operator $\mathcal{O}_n$ in the low-energy EFT, as shown in (\ref{eq_Gn}). Naively, the full correlator $\G$ can then be obtained by summing all $\G_n$ up. 

However, as the order of EFT expansion $n$ grows, the summation in the explicit expression for $\G_n$ in (\ref{eq_Gn}) will be dominated by large $\ell$. Thus, from the saddle point approximation, we can estimate correlators $\G_{n}$ in (\ref{eq_Gn}) for large $n\gg 1$:
\begin{align}
    \G_n\sim\sum_{\ell=L}^{\infty}(-1)^{n+\ell}\ell^{2n}e^{-\ell[\xi(r_1)-\xi(r_2)]}\sim (2n)^{2n}e^{-2n}.
\end{align}
So, the low-energy correlator $\G_n$, as the expansion over $\wt\nu^{-2n}$, shows factorial growth when $n\to \infty$. As a result, the naive sum $\sum\limits_{n=0}^\infty\G_n$ is ill defined. This phenomenon implies that the low-energy correlators we find are just the asymptotic expansion of the UV correlator when $\wt\nu\gg 1$. Nonperturbative effects proportional to $e^{-\pi\wt\nu}$ would kick in and invalidate the EFT description at a certain finite order of the low-energy expansion. 

As we will show below, the spectral representation we introduced above appears useful as the analytic continuation of the divergent sum. Like the familiar Borel transformation, we can recover the full background $\G_{\text{BG}}$ in (\ref{eq_GBG}) after performing the resummation inside the integral representation; Also, we will see that the new ``\emph{on-shell} poles'' that emerge from the resummation give us the CC signals, which correspond to the pair production of massive scalars in the bulk. Together with the dispersive bootstrap, the result here provides a bidirectional reconstruction between the EFT background and the on-shell signal, and thereby provides a resurgence of the UV correlator.  Below, we will carry out this ``spectral resurgence'' and recover nonperturbative effects that cannot be directly seen in the low-energy EFT.

\subsection{Spectral Resummation}
\label{sec_spect_resum}

We start from the spectral representation of the low-energy correlator in (\ref{eq_RepU}). From this expression, we can interpret the sum $\sum\limits_{n=0}^\infty\G_n$ as a sum over all the spectral integrals \emph{and} then switch the order of the sum and the integration:
\begin{align}
    \G(r_1,r_2)\sim \sum_{n=0}^{\infty}\G_n\sim \sum_{\aa=\pm} \ii\aa \int_{-\infty}^{+\infty}\di \mu\,\bigg[\sum_{n=0}^\infty\rho_n(\mu)\bigg]\mathcal{U}^{p_1}_\aa(\mu;r_1)\mathcal{U}^{p_2}_\aa(\mu;r_2).
\end{align}
The sum over all spectral functions $\rho_n$ produces a new spectral density which we call $\rho(\mu)$:
\begin{align}
    \rho(\mu)\equiv \sum_{n=0}^{\infty}\rho_n(\mu)=-\sum_{n=0}^{\infty} \frac{1}{\pi\wt\nu^2}\mu\sinh(\pi\mu)\left(\frac{\mu}{\wt\nu}\right)^{2n}=\frac{\mu\sinh(\pi\mu)/\pi}{\mu^2-\wt\nu^2}.
\end{align}
Clearly, the resummation introduces new poles $\mu=\pm\wt\nu$ of the spectral integrand. These two poles correspond to the intermediate states with the scaling dimensions $\Delta=3/2\pm\ii\wt\nu$, which are nothing but the two modes of the heavy scalar $\sigma$ that we integrated out in Sec.\,\ref{sec_EFT}. We can expect that these ``on-shell poles'' give us exponentially small CC signals as a result of on-shell propagation of the heavy particle. 

At the same time, however, a new subtlety appears here: the poles $\mu=\pm\wt\nu$ are just on the real axis, which is the integration path. Therefore, the above definition only determines the principal value of the density function around these poles. To determine the whole density function, we need a prescription about how to deform the contour to go around these on-shell poles. This additional piece of information cannot be provided by the local EFT itself but should be introduced from a suitable initial condition for the newly emerged degree $\sigma$. As is well known, this initial condition will be translated to an $\ii\epsilon$-prescription, here imposed on the spectral density. In the SK formalism, the $\ii\epsilon$-prescription for SK branches with $\aa=+,-$ is different. With this branch dependence in mind, we can take the following ansatz for the density function:
\begin{align}
\label{eq_rhoUV}
    \rho_\aa (\wt\nu;\mu)=\frac{\mu\sinh(\pi\mu)}{\pi}\left[\frac{\beta(\wt\nu)}{\mu^2-\wt\nu^2+\ii\aa\epsilon}+\frac{1-\beta(\wt\nu)}{\mu^2-\wt\nu^2-\ii\aa\epsilon}\right].
\end{align}
Here, we take care of the difference of $\ii\ep$-prescription for the two SK branches by writing the pole shift as $\pm\ii\aa\ep$. The coefficient $\be(\nu)$ then encodes the boundary condition and could be $\mu$-dependent, whose precise form will be determined below. 

With the above preparation about the density function, we can now assert that the resummed correlator $\G(r_1,r_2)$ has the following spectral representation: 
\begin{align}
    \G(r_1,r_2)=\sum_{\aa=\pm}\ii\aa\int_{-\infty}^{+\infty}\di \mu\,\rho_\aa(\wt\nu;\mu)\mathcal{U}^{p_1}_\aa(\mu;r_1)\mathcal{U}^{p_2}_\aa(\mu;r_2).
\end{align}

We can again perform this integral by closing the integration contour appropriately and collecting residues from all enclosed poles. The newly introduced coefficient $\beta(\wt\nu)$ clearly does not affect the asymptotic behavior of the integrand at large $\mu$, and therefore the analysis is parallel to Sec.\,\ref{sec_compute_contact}. The main step is still to close the contour for each of the four terms with $\cc_1,\cc_2=\pm$ from the side where $\Im (\cc_1\mu)<0$, assuming $r_1<r_2<1$. All poles inside the integration contour are classified in Table \ref{tab_pole} with all $\ell=0,1,2,\cdots$. 
\begin{table}[t]
\centering
\caption{All relevant poles in the $\Im(\cc\mu)<0$ half-plane}
\label{tab_pole}
\vspace{4mm}
\begin{tabular}{|c|cc|}
\hline
Poles                          & \multicolumn{1}{c|}{$\cc_1=\cc_2=\cc$}  & $\cc_1=-\cc_2=\cc$                  \\ \hline
\multirow{2}{*}{Off-shell}     & \multicolumn{2}{c|}{$\mu=-\ii \cc(\ell+1)$}               \\ \cline{2-3} 
                               & \multicolumn{1}{c|}{-}                  & $\mu=-\ii\cc(\ell+p_2+\frac{5}{2})$ \\ \hline
\multicolumn{1}{|l|}{On-shell} & \multicolumn{2}{c|}{$\mu=\pm\wt\nu-\ii\cc\epsilon$}                        \\ \hline
\end{tabular}
\end{table}

\paragraph{EFT background}
We first consider the off-shell contributions. Since the exact cancellation of residues of $\mu=-\ii\cc(\ell+1)$ happens at all orders in the EFT, their total contribution must vanish in UV. So the rest of off-shell poles $\mu=-\ii\cc(\ell+p_2+\fr{5}{2})$ will give the full EFT part of the UV correlator:
\begin{align}
\label{eq_BG1}
    \G_{\text{BG}}(r_1,r_2)=&\sum_{\ell=0}^{\infty}\frac{2(-1)^{\ell}\cos\left(\pi p_{12}/2\right)\Gamma(\ell+p_{12}+5)}{\left[\left(\ell+p_2+5/2\right)^2+\wt\nu^2\right]\ell!}r_1^3\left(\frac{r_1}{r_2}\right)^{\ell+p_2+1}\n\\
    &\times{}_2\mathrm{F}_1\left[\bgm \frac{\ell+p_{12}+5}{2},\frac{\ell+p_{12}+6}{2}\\\ell+p_2+\frac{7}{2}\edm\Bigg|r_1^2\right]{}_2\mathrm{F}_1\left[\bgm -\frac{\ell}{2},-\frac{\ell-1}{2}\\-\ell-p_2-\frac{3}{2}\edm\Bigg|r_2^2\right].
\end{align} 
It can be proved that this expression is identical to the background part of the UV correlator in (\ref{eq_GBG}). We provide the proof in App.\,\ref{app_BG}. Interestingly, after the resummation, the resulting expression for the background depends on our choice of $r_1<r_2$, which is not the case for individual EFT terms $\G_n$. Also, it is not surprising that the background is independent of the choice of $\beta(\wt\nu)$, namely the initial condition for $\sigma$. From the viewpoint of differential equations, the background is the inhomogeneous solution, whose form is completely fixed by the regularity condition at small $r_1$ and small $r_1/r_2$ and is thus independent of the boundary conditions. From the viewpoint of spectral integral, the $\ii\epsilon$-prescription does not play a role if the poles do not come from the spectral density. 

\paragraph{On-shell poles}
Next, we turn to the on-shell poles. As shown in Table \ref{tab_pole}, we should pick up two poles in the $\Im(\cc\mu)<0$ half-plane at $\mu=\pm\wt\nu-\ii\cc\epsilon$. Since these poles come from the spectral density $\rho_\aa(\wt\nu;\mu)$, their residues will depend on $\beta(\wt\nu)$: The residue of $\mu=\cc(\aa\wt\nu-\ii\epsilon)$ is proportional to $\beta(\wt\nu)$ while the residue of $\mu=\cc(-\aa\wt\nu-\ii\epsilon)$ is proportional to $[1-\beta(\wt\nu)]$. Then, the contribution from $\cc_1=\pm\cc_2$ will give us two kinds of contributions $\G_{\text{on-shell}}^{(\pm)}$: 
\begin{align}
\label{eq_GOnShell}
    &\G_{\text{on-shell}}^{(\pm)}(r_1,r_2)=\sum_{\aa=\pm}\G_{\text{on-shell}}^{(\pm)\aa}(r_1,r_2).
\end{align}
Here we have kept the SK index $\aa$ explicit for later analytic continuation. Their explicit expressions are:
\begin{align}
\label{eq_NSpart}
  \G_{\text{on-shell}}^{(\pm)\aa}(r_1,r_2)=&~-\frac{4}{\pi}e^{-\ii\aa(p_{12}+5)\pi/2}\sinh (\pi\wt\nu)\n\\
  &\times\Big[\be\mathcal{Z}(\aa\wt\nu,p_1;r_1)\mathcal{Z}(\pm\aa\wt\nu,p_2;r_2)-(1-\be)\mathcal{Z}(-\aa\wt\nu,p_1;r_1)\mathcal{Z}(\mp\aa\wt\nu,p_2;r_2)\Big],
\end{align}
where we have defined a function $\mathcal{Z}(\wt\nu,p;r)$ for convenience: 
\bge
\label{eq_Z}
\mathcal{Z}(\wt\nu,p;r)=\Gamma\big[-\ii\wt\nu,p+\ii\wt\nu+\fr52\big]\Big(\FR{r}{2}\Big)^{\ii\wt\nu+3/2}{}_2\mathrm{F}_1\left[\bgm \frac{p+\ii\wt\nu}{2}+\frac{5}{4},\frac{p+\ii\wt\nu}{2}+\frac{7}{4}\\1+\ii\wt\nu\edm\Bigg|r^2\right].
\ede 
Clearly, these two pieces have the form of CC signals of a single scalar exchange of mass parameter $\wt\nu$ in the UV: $\G_{\text{on-shell}}^{(+)}$ has the form of the nonlocal signal and $\G_{\text{on-shell}}^{(-)}$ the local signal. However, they are not the complete CC signals due to the missing of certain boundary information. So, the conclusion of this subsection is that we can recover some but not all exponentially small effects in the UV, and, without imposing further boundary conditions, the final result of our resummation is:
\bge
\label{eq_Gresum}
\G_\text{BG}+\G_\text{on-shell}^{(+)}+\G_\text{on-shell}^{(-)},
\ede
where $\G_\text{BG}$ is given in (\ref{eq_BG1}) and is identical to the background of the UV correlator, and $\G_\text{on-shell}^{(\pm)}$ are given in (\ref{eq_NSpart}) and are only part of the CC signals of the UV correlator. 

\subsection{Boundary conditions}
\label{sec_bdry_cond}

Although the resummation performed above recovers some ``signals'' that have exponential dependence $\sim e^{-\pi\wt\nu}$ in the large mass limit, the result is not complete. The underlying physical reason is that, by removing a propagating degree in the low-energy EFT expansion, we have also removed information about its boundary conditions. Without providing appropriate boundary conditions, it is impossible to uniquely recover the original correlator in the UV theory in which the heavy particle $\si$ was assumed to be created from a BD initial state. 

Below, we will show that the two missing conditions must be imposed to recover the correct UV correlator: One is the final-state unitarity, and the other is the BD initial condition. 

\paragraph{Unitarity}
The unitarity is best understood with bulk time integrals: When computing the UV correlator following the time integral in (\ref{eq_GTimeInt}), we have four SK branches labeled by $\aa_{1},\aa_2=\pm$. However, when working with the low-energy EFT, only two SK branches survive, labeled by $\aa=\pm$ in (\ref{eq_GnTimeInt}). Clearly, the surviving two branches correspond to $\aa=\aa_1=\aa_2$. So, the EFT calculation only keeps track of $(\pm\pm)$ branches of the UV integral and is ignorant about the $(\pm\mp)$ branches, a phenomenon also highlighted in \cite{Green:2024cmx}. 

However, in the UV theory, the existence of $(\pm\mp)$ is enforced by the completeness of final states. Recall that in the SK formalism, we trace out final states by inserting a complete basis $1=\sum |\text{out}\ra\la\text{out}|$ at the final time \cite{Chen:2017ryl}, which leads to the opposite-sign propagators $\wt D_{\pm\mp}$ for any bulk field. In particular, they together with the same sign propagators $\wt D_{\pm\pm}$ satisfy the relation:
\begin{align}
\label{eq_4D}
    \wt D_{\pm\pm}(z_1,z_2)=\wt D_{\mp\pm}(z_1,z_2)\theta(z_1-z_2)+\wt D_{\pm\mp}(z_1,z_2)\theta(z_2-z_1).
\end{align}
From the bulk perspective, the cutting rule of CC signals \cite{Tong:2021wai, Liu:2024str} tells us that the signals in the $(\pm\pm)$ branches can be computed from the time integral below:
\begin{align}
    &\G_{\text{on-shell}}^{(+)\aa}(r_1,r_2)+\G_{\text{on-shell}}^{(-)\aa}(r_1,r_2)=-\int_{-\infty(1-\ii\aa\epsilon)}^{0}\di z_1\,\di z_2\,(-z_1)^{p_1} (-z_2)^{p_2} e^{\ii\aa z_{12}}\wt D_{(-\aa)\aa}(r_{1}z_{1},r_{2}z_{2}).
\end{align}
Meanwhile the bulk integral for correlators in the $(\pm\mp)$ branches has a similar form, which can be directly captured from the above $(\pm\pm)$ contributions through analytic continuation:
\begin{align}
\label{eq_AC}
    \G_{(-\aa)\aa}(r_1,r_2)=&\int_{-\infty(1+\ii\aa\epsilon)}^{0}\di z_1\,\int_{-\infty(1-\ii\aa\epsilon)}^{0}\di z_2\,(-z_1)^{p_1} (-z_2)^{p_2} e^{\ii\aa(-z_1+z_2)}\wt D_{(-\aa)\aa}(r_{1}z_{1},r_{2}z_{2})\n\\
    =&~e^{\ii\aa \pi p_{1}}\Big[\G_{\text{on-shell}}^{(+)\aa}(r_1 e^{\ii\aa\pi},r_2)+\G_{\text{on-shell}}^{(-)\aa}(r_1e^{\ii\aa\pi},r_2)\Big].
\end{align}
Here, the phase of $-r_1$ after analytic continuation $r_1\to e^{\ii\aa\pi}r_1$ depends on the SK index $\aa$. A more detailed discussion on the analytic continuation of these bulk time integrals can be found in \cite{Liu:2024xyi}. As a result, the final-state unitarity demands that we should replace the previously computed signals in (\ref{eq_GOnShell}) as:
\begin{align}
\label{eq_Gunitarized}
  \G_{\text{on-shell}}^{(\pm)\aa}(r_1,r_2)\to  \G_{\text{unitary}}^{(\pm)\aa}(r_1,r_2)\equiv \G_{\text{on-shell}}^{(\pm)\aa}(r_1,r_2)+e^{\ii\aa\pi p_1}\G_{\text{on-shell}}^{(\pm)\aa}(r_1 e^{\ii\aa\pi},r_2) .
\end{align}
With this replacement, we find the ``unitarized'' on-shell pieces $\G_\text{unitary}^{(\pm)}\equiv\sum\limits_{\aa=\pm}\G_\text{unitary}^{(\pm)\aa}$ as:
\begin{align}
  \G_\text{unitary}^{(\pm)}\equiv &
    -\sum_{\aa=\pm}\frac{4}{\pi}e^{-\ii\aa(p_{12}+5)\pi/2}\sinh (\pi\wt\nu)\Big[\beta(1+e^{-\pi\wt\nu+\ii\aa\pi(p_1+3/2)})\mathcal{Z}(\aa\wt\nu,p_1;r_1)\mathcal{Z}(\pm\aa\wt\nu,p_2;r_2)
   \n\\
   &-(1-\be)(1+e^{\pi\wt\nu+\ii\aa\pi(p_1+3/2)}) \mathcal{Z}(-\aa\wt\nu,p_1;r_1)\mathcal{Z}(\mp\aa\wt\nu,p_2;r_2)\Big].
\end{align}
In summary, after imposing the unitary condition of final states, the resummed correlator is upgraded from (\ref{eq_Gresum}) to:
\bge
\label{eq_Gunitary}
\G_\text{BG}+\G_\text{unitary}^{(+)}+\G_\text{unitary}^{(-)}.
\ede
This represents the correlator computed from a consistent SK path integral by tracing out a complete basis of final states. However, there is still a free parameter $\be(\wt\nu)$ in the above expression, to be determined by an initial condition, which we discuss next.

\paragraph{Bunch-Davies State} Since the unitary condition was imposed on the final-state tracing procedure in the above discussion, the result (\ref{eq_Gunitary}) is still ignorant about the initial condition for the heavy state. In fact, it represents a correlator of single-massive exchange where the massive particle is created from one of many dS-invariant initial (pure or mixed) states parameterized by $\beta(\wt\nu)$. The way to fix this $\be$-dependence is well known: The correct BD initial condition leads to cancellation of spurious poles in the folded limit when either $r_1\to 1$ or $r_2\to 1$. It also leads to correct factorization of the correlator in the partial-energy limit where $r_1\to -1$ or $r_2\to -1$. We can use either of these two limits to fix $\be$, but we consider both for completeness. Different limiting forms of $\mathcal{Z}(\wt\nu,p;r)$ are collected in App.\,\ref{app_2F1}.

\paragraph{(i) Folded limit} It is well known that a correlator generated from BD initial condition should be regular in the folded limit when either $r_1\to 1$ or $r_2\to 1$ \cite{Arkani-Hamed:2018kmz}. For simplicity, let us consider the limit $r_1\to 1$. However, there is a small technical subtlety: Since we always consider $|r_1|<|r_2|$ in this work, sending $r_1\to 1$ means that we need to take $|r_2|>1$ which is beyond the physical region. However, this poses no challenge to us since all our expressions stay well within the validity range even when $|r_2|>1$ as long as $|r_1|<|r_2|$. With this point in mind, we can now check the limit $r_1\to 1$. It turns out that the $(\pm\pm)$ are automatically regular in this limit for general $\be$. So, we only need to check the $(\mp\pm)$ branches. Taking the $\aa_1=-\aa_2=-$ branch for example, we can use (\ref{eq_AC}) and check the folded limit as follows:
\begin{align}
   \lim_{r_1\to 1} \G_{-+}(r_1,r_2) =&\lim_{r_1\to 1} e^{\ii\pi p_1}\Big[\G_{\text{on-shell}}^{(+)\aa}(r_1 e^{\ii\aa\pi},r_2)+\G_{\text{on-shell}}^{(-)\aa}(r_1e^{\ii\aa\pi},r_2)\Big]\n\\
   =&~\big[\beta e^{-\pi\wt\nu}+(1-\beta)e^{\pi\wt\nu}\big]\frac{2^{p_1+2}\ii\pi^{-1/2}e^{\ii\pi(p_1-p_2)/2}\Gamma(p_1+2)}{(1-r_1^2)^{p_1+2}} \n\\
    &\times\big[\mathcal{Z}(\wt\nu,p_2;r_2)+\mathcal{Z}(-\wt\nu,p_2;r_2)\big]+\text{regular terms}.
\end{align}
So, the absence of a singularity demands $\beta e^{-\pi\wt\nu}+(1-\beta) e^{\pi\wt\nu}=0$, which determines the value of $\be$:
\begin{align}
\label{eq_beta}
    \beta(\wt\nu) = \frac{e^{\pi\wt\nu}}{2\sinh(\pi\wt\nu)}.
\end{align}
This is also the correct coefficient for the spectral representation of both the massive propagator and correlators discussed in \cite{Melville:2024ove,Werth:2024mjg}. 
Now, it is straightforward to check that the unitary correlator (\ref{eq_Gunitary}) with the above value of $\be$ is identical to the UV correlator in (\ref{eq_Gresult}). In particular, the $\G_\text{unitary}^{(+)}$ is identical to the nonlocal signal in (\ref{eq_GNS}) and $\G_\text{unitary}^{(-)}$ to the local signal in (\ref{eq_GLS}). So, we have demonstrated that the EFT resummation together with the unitarity and BD initial conditions can correctly reproduce the UV correlator with a propagating intermediate heavy particle.

\paragraph{(ii) Partial-energy limit}
The $\be$ parameter can also be fixed by requiring consistent factorization in the partial-energy limit ($r_1\to -1$ or $r_2\to -1$), where the correlator is factorized into the product of a singular branch cut, a flat-space amplitude (which is a constant in our case), and a shifted three-point function with BD initial condition. This factorization has also been extensively studied in the literature for both dS-invariant and more general cases \cite{Arkani-Hamed:2018kmz,Baumann:2020dch,Baumann:2021fxj,Xianyu:2025lbk}. In our case, it is convenient to consider the limit $r_2\to -1$, where the shifted three-point function can be generated from $\mathcal{U}^p_\aa(r)$ in (\ref{eq_3pt}) and the correct partial-energy limit reads:
\begin{align}
\label{eq_PE1}
    \lim_{r_2\to -1} \G(r_1,r_2) =&~ -\FR{2^{p_2+3/2}\Gamma(p_2+2)}{(1-r_2^2)^{p_2+2}}\n\\
    &\times\sum_{\aa=\pm}e^{-\ii\aa(p_2+5/2)\pi/2}\Big[\mathcal{U}_{\aa}^{p_1}(e^{-\ii\aa\pi}r_1)+e^{\ii\aa p_1\pi}\mathcal{U}_{\aa}^{p_1}(r_1)\Big]+\text{regular terms}.
\end{align}
Here, to avoid the ambiguity caused by the branch cut on the negative real axis of $r_2$-plane, we define the partial-energy limit $r_2\to -1$ of $\G(r_1,r_2)$ as:
\begin{align}
    \lim_{r_2\to -1} \G(r_1,r_2) \equiv \FR12\lim_{r_2\to 1}\Big[\G(r_1,r_2e^{\ii\pi})+\G(r_1,r_2e^{-\ii\pi})\Big].
\end{align}
The background $\G_\text{BG}$ is regular in the limit $r_2\to -1$ which can be seen through either a standard Landau analysis or a direct computation. So, to isolate the singular contributions, we only need to focus on the on-shell contributions found above:
\begin{align}
\label{eq_PE2}
    &\lim_{r_2\to -1} \G(r_1,r_2) = \lim_{r_2\to -1}\Big[ \G_\text{unitary}^{(+)}(r_1,r_2)+\G_\text{unitary}^{(-)}(r_1,r_2)\Big]\n\\
    =&~\sum_{\aa=\pm}\FR{2^{p_2+3}\pi^{-1/2}e^{-\ii\aa(p_{12}+5)\pi/2}\Gamma(p_2+2)\sinh(\pi\wt\nu)}{(1-r_2^2)^{p_2+2}}\Big[\beta\big(1+e^{-\pi\wt\nu+\ii\aa(2p_1+3)\pi/2}\big) \mathcal{Z}(\aa\wt\nu,p_1;r_1)
   \n\\
    &  -(1-\beta)\times \big(1+e^{\pi\wt\nu+\ii\aa(2p_1+3)\pi/2}\big) \mathcal{Z}(-\aa\wt\nu,p_1;r_1)\Big]+\text{regular terms}.
\end{align}
It is now easy to check that taking the value in (\ref{eq_beta}) for $\be$ makes the above singular term approach the correct partial-energy limit. So, once again, the BD initial condition can be as well imposed by requiring the correct partial-energy limit.

\section{Resurgence with Borel Resummation}
\label{sec_Borel}

In the previous three sections, we have achieved the resummation of EFT expansions in two ways: either through the resummation of boundary differential operators in Sec.\;\ref{sec_boundary}, or through the spectral resummation in Sec.\;\ref{sec_signal}. However, the resurgence is more often discussed together with the Borel resummation. In this section, we revisit our problem with the more standard Borel approach. This discussion provides a useful complement to the spectral resummation in Sec.\,\ref{sec_signal}: Instead of resumming the spectral density first, here we directly Borel resum the large-mass expansion of the EFT correlator. Our starting point is still the EFT correlator at the $n$-th order in (\ref{eq_Gn}):
\begin{align}
    \G_n=&\frac{4\cos\left(\pi p_{12}/2\right)r_1r_2}{\wt\nu(1-r_1^2)^{p_1/2+1}(1-r_2^2)^{p_2/2+1}}\sum_{\ell=0}^{\infty}\frac{(-1)^{n+\ell}}{\ell!}\left(\frac{\ell+p_{2}+5/2}{\wt\nu}\right)^{2n+1}\n\\
    &\times\Gamma(\ell+2p_2+5)\text{Q}^{p_1+2}_{\ell+p_2+2}(r_1^{-1})\text{P}^{-p_2-2}_{\ell+p_2+2}(r_2^{-1}).
\end{align}
Here, for later convenience, we have expressed the hypergeometric functions in (\ref{eq_Gn}) in terms of Legendre functions $\text{P}^{\al}_{\be}$ and $\text{Q}^{\al}_{\be}$ defined in App.\,\ref{app_2F1}. 

We have seen from Sec.\,\ref{sec_signal} that the large-order growth of $\G_n$ is factorial, namely, $\G_n\sim (2n)^{2n}e^{-2n}\sim (2n)!$ for large $n$. This makes the Borel transform immediate: From the elementary integral
\begin{align}
    \int_{0}^{\infty}\di s\, e^{-s\wt\nu} s^{2n}=\wt\nu^{-2n-1}(2n)!,
\end{align}
we can write the Borel resummation of $\sum_n\G_n$ as
\begin{align}
\label{eq_BorelTransform}
    \G_\text{Borel}(r_1,r_2)=\frac{2\cos\left(\pi p_{12}/2\right)r_1r_2}{\wt\nu(1-r_1^2)^{p_1/2+1}(1-r_2^2)^{p_2/2+1}}\int_{0}^{\infty}\di s\, e^{-s\wt\nu}B(r_1,r_2;s),
\end{align}
with the Borel transform of the summand being:
\begin{align}
\label{eq_BorelIntegrand}
    B(r_1,r_2;s)=&\sum_{\ell=0}^{\infty}\sum_{n=0}^{\infty}\frac{2(-1)^{n+\ell}s^{2n}\left(\ell+p_2+5/2\right)^{2n+1}}{(2n)!\ell!}\Gamma(\ell+2p_2+5)\text{Q}^{p_1+2}_{\ell+p_2+2}(r_1^{-1})\text{P}^{-p_2-2}_{\ell+p_2+2}(r_2^{-1})\n\\
    =&\sum_{\cc=\pm}\sum_{\ell=0}^{\infty}e^{\ii \cc s(\ell+p_2+5/2)}\frac{(-1)^\ell}{\ell!}(\ell+p_2+5/2)\Gamma(\ell+2p_2+5)\text{Q}^{p_1+2}_{\ell+p_2+2}(r_1^{-1})\text{P}^{-p_2-2}_{\ell+p_2+2}(r_2^{-1})\n\\
    =&\sum_{\cc=\pm}e^{\ii\cc s(p_{2}+5/2)}\mathcal{B}(r_1,r_2;-e^{\ii\cc s}).
\end{align}
At this point, we can try to finish the integral (\ref{eq_BorelTransform}) to get a resummed result for $\G$. However, as is well known, we have a remaining freedom of choosing the direction of the contour: Typically, if we let the contour approach $\infty$ from different directions on the complex plane, the result generally reproduces the same asymptotic series $\sum_n\G_n$. However, as is well known, when deforming the contour past singularities of the integrand $B(r_1,r_2;s)$, we will pick up additional nonperturbative contributions which are exponentially small in $\wt\nu$ in the large $\wt\nu$ limit. These are exactly the CC signals. So, to detect CC signals through resummation, we need to study the singularity structure of the Borel transform $B(r_1,r_2;s)$. For our purpose, it is enough to consider the following generating function: 
\begin{align}
    \label{eq_B}
    \mathcal{B}(r_1,r_2;w)=&\sum_{\ell=0}^{\infty}\frac{(\ell+p_2+5/2)}{\ell!}\Gamma(\ell+2p_2+5)\text{Q}^{p_1+2}_{\ell+p_2+2}(r_1^{-1})\text{P}^{-p_2-2}_{\ell+p_2+2}(r_2^{-1})w^\ell.
\end{align}
Below, we perform the analysis in the complex $w$-plane in Sec.\;\ref{sec_borel_singularity}, and later in Sec.\;\ref{sec_borel_ambiguities} we will translate the result back to $s$-plane and use it to understand the resummed correlator. 

\subsection{Singularities of the Borel Transform}
\label{sec_borel_singularity}

In this subsection we study the singularities of $\mathcal{B}(r_1,r_2;w)$. The upshot of this subsection is that $\mathcal{B}(r_1,r_2;w)$ is divided into two parts with different analytic properties:
\begin{align}
  \mathcal{B}(r_1,r_2;w)=\sum_{\dd=\pm}\mathcal{B}_{\dd}(r_1,r_2;w);
\end{align}
\begin{align}
\label{eq_B1}
  \mathcal{B}_{+}(r_1,r_2;w)=&\sum_{\ell=0}^{\infty}(\ell+p_2+5/2)\text{Q}^{p_1+2}_{\ell+p_2+2}(r_1^{-1})\text{P}^{p_2+2}_{\ell+p_2+2}(r_2^{-1})w^\ell,\\
\label{eq_B2}
  \mathcal{B}_{-}(r_1,r_2;w)=&-\frac{2\sin(\pi p_2)}{\pi}\sum_{\ell=0}^{\infty}(\ell+p_2+5/2)\text{Q}^{p_1+2}_{\ell+p_2+2}(r_1^{-1})\text{Q}^{p_2+2}_{\ell+p_2+2}(r_2^{-1})w^\ell.
\end{align}
To derive these expressions, we use the connection formula of associated Legendre functions in (\ref{eq_PQconnection}). As always, we work with a particular physical region $r_1<r_2<1$. The above expressions for $\mathcal{B}_{\pm}$ are Taylor series around $w=0$ with finite convergence regions, showing that $\mathcal{B}_{\pm}$ are analytic near $w=0$ but their analytic properties beyond the convergence regions are not obvious. To make progress, we use the integral representations for the associated Legendre functions:
\begin{align}
\label{eq_Qint}
  \text{Q}^{p_i+2}_{\lambda-1/2}(r_i^{-1})=&~\sqrt{\frac{\pi}{2}}\frac{\sinh^{p_i+2}\xi(r_i)}{\Gamma(-p_i-3/2)}e^{-\lambda\xi(r_i)}\int_0^{\infty}\di t_i\,\frac{e^{-\lambda t_i}}{[\cosh(\xi(r_i)+t_i)-r_i^{-1}]^{p_i+5/2}},\quad (i=1,2)\\
\label{eq_Pint}
  \text{P}^{p_2+2}_{\lambda-1/2}(r_2^{-1})=&~\frac{1}{\sqrt{2\pi}}\frac{\sinh^{p_2+2}\xi(r_2)}{\Gamma(-p_2-3/2)}e^{\lambda\xi(r_2)}\int_0^{2\xi(r_2)}\di t_2\,\frac{e^{-\lambda t_2}}{[r_2^{-1}-\cosh(\xi(r_2)-t_2)]^{p_2+5/2}}.
\end{align}
Detailed derivations of these representations are collected in App.\,\ref{app_2F1}. Inserting these representations into (\ref{eq_B1}) and (\ref{eq_B2}), and finishing the $\ell$-summation, we get the analytic continuation of Taylor series for $\mathcal{B}_{\pm}(r_1,r_2;w)$ as the following integral representations: 
\begin{align}
  \mathcal{B}_{\pm}(r_1,r_2;w)=&\int_0^{\infty}\di t\,e^{-(p_2+5/2)t}R_{\pm}(r_1,r_2;t)\sum_{\ell=0}^{\infty}(\ell+p_2+5/2)e^{-\ell[\xi(r_1)\mp\xi(r_2)+t]}w^\ell\n\\
  =&\int_0^{\infty}\di t\,e^{-(p_2+5/2)t}R_{\pm}(r_1,r_2;t)\frac{p_2+5/2-\left(p_2+3/2\right)e^{-\xi(r_1)\pm\xi(r_2)-t}w}{(1-e^{-\xi(r_1)\pm\xi(r_2)-t}w)^2}.
\end{align}
Here, the functions $R_{\pm}(r_1,r_2;t)$ are defined by
\begin{align}
  &R_{+}(r_1,r_2;t)=\frac{\sinh^{p_1+2}\xi(r_1)\sinh^{p_2+2}\xi(r_2)e^{-(p_2+5/2)[\xi(r_1)-\xi(r_2)]}}{2\Gamma(-p_1-3/2)\Gamma(-p_2-3/2)}\n\\
  &\times\int_{0}^{\min\{t,2\xi(r_2)\}}\frac{\di t_2}{[\cosh(\xi(r_1)+t-t_2)-r_1^{-1}]^{p_1+5/2}[r_2^{-1}-\cosh(\xi(r_2)-t_2)]^{p_2+5/2}},\\
  &R_{-}(r_1,r_2;t)=-\frac{\sin(\pi p_2)\sinh^{p_1+2}\xi(r_1)\sinh^{p_2+2}\xi(r_2)e^{-(p_2+5/2)[\xi(r_1)+\xi(r_2)]}}{\Gamma(-p_1-3/2)\Gamma(-p_2-3/2)}\n\\
  &\times\int_{0}^{t}\frac{\di t_2}{[\cosh(\xi(r_1)+t-t_2)-r_1^{-1}]^{p_1+5/2}[\cosh(\xi(r_2)+t_2)-r_2^{-1}]^{p_2+5/2}},
\end{align}
and are independent of $w$. We further use integration by parts to recast the above integral representations into:
\begin{align}
  \mathcal{B}_{\pm}(r_1,r_2;w)=&\frac{R_{\pm}(r_1,r_2;0)}{1-e^{-\xi(r_1)\pm\xi(r_2)}w}+\int_0^{\infty}\di t\,\frac{e^{-(p_2+5/2)t}}{1-e^{-\xi(r_1)\pm\xi(r_2)-t}w}\partial_tR_{\pm}(r_1,r_2;t).
\end{align}

Then, it is clear that the first term above produces an isolated pole at $w=e^{\xi(r_1)\mp\xi(r_2)}$. The remaining terms are $t$-integrals, whose integrand contains a pole at $w=e^{t+\xi(r_1)\mp\xi(r_2)}$. After integrating over $t\in(0,\infty)$, they produce a branch cut emanating from $w=e^{\xi(r_1)\mp\xi(r_2)}$ towards infinity in the positive real direction. As we will see in the next subsection, picking up these residues properly will give us both the local signal and the nonlocal signal. 

\subsection{Borel integral and ambiguities}
\label{sec_borel_ambiguities}

\begin{figure}[t]
\centering
\includegraphics[width=0.55\textwidth]{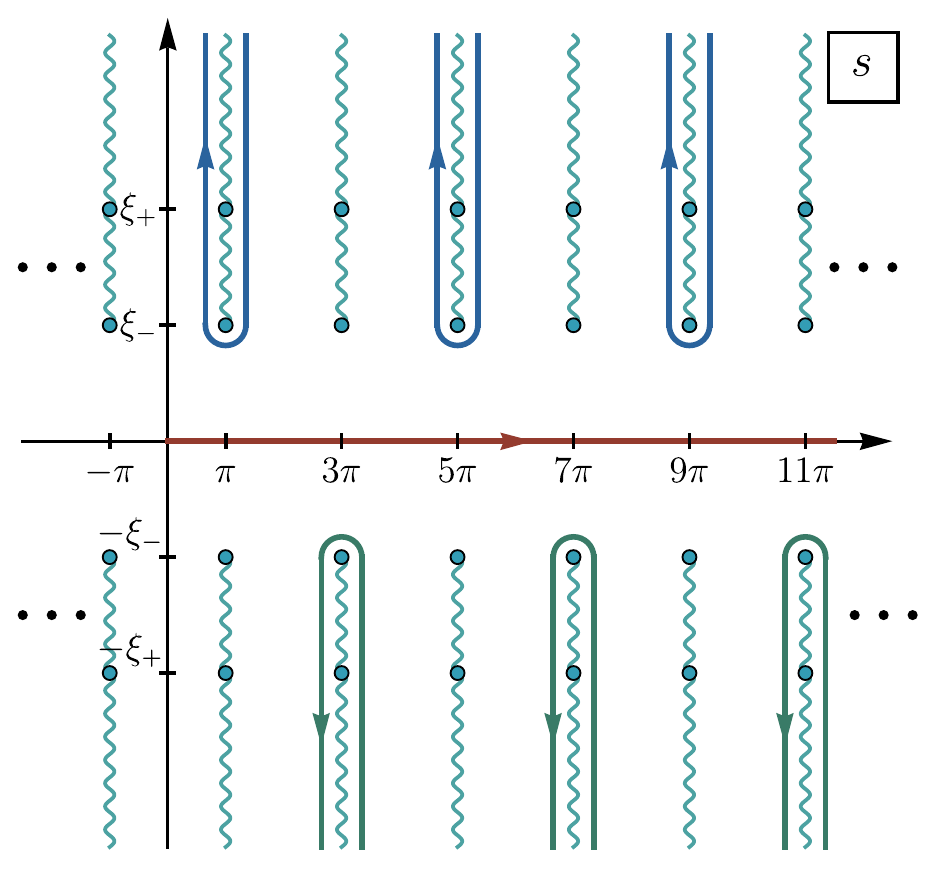}
\caption{The analytic structure of the Borel transform $B(r_1,r_2;s)$ in (\ref{eq_BorelIntegrand}) on the complex $s$ plane. The cyan dots and wavy lines show the poles and cuts from (\ref{eq_BorelSingularities}), with $\xi_\pm\equiv \xi(r_1)\pm\xi(r_2)$. The red contour on the positive real axis shows the original Borel integral that gives rise to the background. For $p_1,p_2\in\mathbb{Z}$ and $\aa=+$, the blue (upper-half plane, $\cc=-$) and green (lower-half plane, $\cc=+$) contours show the extra contributions to the Borel integral respecting the BD initial condition which give rise to correct on-shell contributions. }
\label{fig_BorelPlane}
\end{figure}

Above we have analyzed the analytic properties of Borel function $\mathcal{B}$ on the complex $w$-plane and determined all its singularities. Now we come back to the $s$-plane. With the map $w=-e^{\ii \cc s}$, they are mapped to periodically repeating branch cuts of $\mathcal{B}_{\pm}(r_1,r_2;-e^{\ii\cc s})$:
\begin{align}
\label{eq_BorelSingularities}
    &s=(2m+1)\pi-\ii\cc\big[\xi(r_1)\mp\xi(r_2)+t\big],&&(m\in\mathbb{Z}, t\in\mathbb{R}_+)
\end{align}
We show these poles and cuts in Fig.\;\ref{fig_BorelPlane}. 
Since the contour of Borel integral (\ref{eq_BorelTransform}) is not determined from EFT alone, the presence of these singularities introduces possibilities/ambiguities of new nonperturbative contributions to the result. In this subsection, we use these observations to determine different contributions to the final resummed correlator. We will see that the background part, which is analytic in the large $\wt\nu$ limit, can be robustly computed. Meanwhile, we can also detect exponentially small local and nonlocal signals by collecting contributions from the residues/discontinuities of the poles/cuts, but their precise forms are subject to ambiguities that should be resolved by proper boundary conditions. 

\paragraph{Background}
The pole structure summarized above shows that there are no divergent poles along the positive real axis. So, we can directly finish the Borel integral along the positive real axis with no obstructions, as shown by the red contour in Fig.\;\ref{fig_BorelPlane}. Indeed, inserting the series (\ref{eq_B}) directly into the Borel integral (\ref{eq_BorelTransform}) gives:
\begin{align}
    \G_{\text{BG}}=&~\frac{2\cos\left(\pi p_{12}/2\right)r_1r_2}{\wt\nu(1-r_1^2)^{p_1/2+1}(1-r_2^2)^{p_2/2+1}}\sum_{\cc=\pm}\int_{0}^{\infty}\di s\,e^{-s\wt\nu}e^{\ii\cc s(p_2+5/2)}\mathcal{B}(r_1,r_2;-e^{\ii\cc s})\n\\
    =&~\frac{4\cos\left(\pi p_{12}/2\right)r_1r_2}{(1-r_1^2)^{p_1/2+1}(1-r_2^2)^{p_2/2+1}}\n\\
    &\times\sum_{\ell=0}^{\infty}\frac{(-1)^\ell}{\ell!}\frac{(\ell+p_2+5/2)\Gamma(\ell+2p_2+5)}{(\ell+p_2+5/2)^2+\wt\nu^2}\text{Q}^{p_1+2}_{\ell+p_2+2}(r_1^{-1})\text{P}^{-p_2-2}_{\ell+p_2+2}(r_2^{-1}),
\end{align}
which is precisely the background in (\ref{eq_BG1}). As mentioned before, deforming the contour across the singularities, we may pick up additional contributions from residues and cuts. However, they are always exponentially suppressed in $\wt\nu$ and thus do not introduce power-suppressed terms to $\G_\text{BG}$. In this sense, the background computed here is insensitive to contour deformation.  

\paragraph{CC signals} Now let us consider contributions from singularities in (\ref{eq_BorelSingularities}). Obviously, there is much freedom to deform the original integration contour along the real axis. Accordingly, the branch cuts in (\ref{eq_BorelSingularities}), labeled by $m\in\mathbb{Z}$, can be either surrounded (in either direction) or not surrounded by the contour. Since we always consider the case $\wt\nu>0$, we always deform the contour such that it remains in the right half-plane with $\Re(s)>0$. So, only poles and cuts with $m=0,1,2,\cdots$ are relevant to our discussion, i.e., they can in principle be wound by the contour. Similar to that we choose different $\ii\epsilon$-prescriptions for different SK branches with $\aa=+,-$ in Sec.\, \ref{sec_signal}, here we also need to choose different integration contours for $B^{+}(r_1,r_2;s)$ and $B^{-}(r_1,r_2;s)$, defined as:
\begin{align}
    B^{\aa}(r_1,r_2;s)=\frac{e^{-\ii\aa \pi p_{12}/2}}{2\cos\left(\pi p_{12}/2\right)}B(r_1,r_2;s).
\end{align}
To parameterize such freedom, we use $b^{\aa}_{\cc,m}\in\mathbb{Z}$ to denote the winding number around the $m$-th branch cuts. Then, collecting the residues and discontinuities from these poles and the corresponding cuts gives:
\begin{align}
    \G_{\text{on-shell}}^{(\cc)\aa}=&-\frac{2\pi\cc e^{-\ii\aa\pi p_{12}/2}r_1r_2}{\wt\nu(1-r_1^2)^{p_1/2+1}(1-r_2^2)^{p_2/2+1}}\sum_{m=0}^{\infty}b^{\aa}_{\cc,m}e^{-(2m+1)\pi\wt\nu}e^{\ii\cc(2m+1)\pi(p_2+5/2)}\n\\
    &\times\sum_{\dd=\pm}e^{(\ii\cc\wt\nu+p_2+5/2)[\xi(r_1)-\dd\xi(r_2)]}\left[R_{\dd}(r_1,r_2;0)+\int_0^\infty\di t\,e^{\ii\cc\wt\nu t}\partial_tR_{\dd}(r_1,r_2;t)\right]\n\\
    =&~\frac{2\pi\ii e^{-\ii\aa\pi p_{12}/2}r_1r_2}{(1-r_1^2)^{p_1/2+1}(1-r_2^2)^{p_2/2+1}}\sum_{m=0}^{\infty}b^{\aa}_{\cc,m}e^{-(2m+1)\pi\wt\nu}e^{\ii\cc(2m+1)\pi(p_2+5/2)}\n\\
    &\times\sum_{\dd=\pm}e^{(\ii\cc\wt\nu+p_2+5/2)[\xi(r_1)-\dd\xi(r_2)]}\int_0^\infty\di t\,e^{\ii\cc\wt\nu t}R_{\dd}(r_1,r_2;t).
\end{align}
The remaining $t$-integrals can be recognized as the integral representations of associated Legendre functions in (\ref{eq_Qint}) and (\ref{eq_Pint}):
\begin{align}
    &e^{(\ii\cc\wt\nu+p_2+5/2)[\xi(r_1)-\xi(r_2)]}\int_0^\infty\di t\,e^{\ii\cc\wt\nu t}R_{+}(r_1,r_2;t)=\text{Q}^{p_1+2}_{-\ii\cc\wt\nu-1/2}(r_1^{-1})\text{P}^{p_2+2}_{-\ii\cc\wt\nu-1/2}(r_2^{-1}),\\
    &e^{(\ii\cc\wt\nu+p_2+5/2)[\xi(r_1)+\xi(r_2)]}\int_0^\infty\di t\,e^{\ii\cc\wt\nu t}R_{-}(r_1,r_2;t)=-\frac{2\sin(\pi p_2)}{\pi}\text{Q}^{p_1+2}_{-\ii\cc\wt\nu-1/2}(r_1^{-1})\text{Q}^{p_2+2}_{-\ii\cc\wt\nu-1/2}(r_2^{-1}).
\end{align}
Therefore, the resummed result can be written as:
\begin{align}
    \G_{\text{on-shell}}^{(\cc)\aa}=&\frac{2\pi\ii e^{-\ii\aa\pi p_{12}/2}r_1r_2}{(1-r_1^2)^{p_1/2+1}(1-r_2^2)^{p_2/2+1}}\left[\sum_{m=0}^{\infty}b^{\aa}_{\cc,m}e^{-(2m+1)\pi\wt\nu}e^{\ii\cc(2m+1)\pi(p_2+5/2)}\right]\n\\
    &\times\Gamma\left[\bgm p_2+5/2-\ii\cc\wt\nu\\-p_2-3/2-\ii\cc\wt\nu\edm\right]\text{Q}^{p_1+2}_{-\ii\cc\wt\nu-1/2}(r_1^{-1})\text{P}^{-p_2-2}_{-\ii\cc\wt\nu-1/2}(r_2^{-1}).
\end{align}

Again, we used the connection formula (\ref{eq_PQconnection}) here. Thus, the new contributions from Borel singularities give rise to the nonperturbative and oscillatory structure of nonlocal and local signals, but with ambiguities parameterized by $b^{\aa}_{\cc,m}$. We have seen from previous sections that $b^{\aa}_{\cc,m}$ is to be determined by boundary conditions rather than the EFT expansion itself. We will not repeat the analysis of boundary conditions given in Sec.\,\ref{sec_signal}. Instead, we directly fix $b^{\aa}_{\cc, m}$ by requiring the same boundary conditions as before. For this purpose, it is easier to work with integer twists $p_1,p_2\in\mathbb{Z}$. Then, $\G_{\text{on-shell}}^{(\cc)\aa}$ satisfying the BD condition can be uniquely determined by taking the following winding numbers:
\begin{align}
  &b^{\aa}_{\aa,2k}=0,&&b^{\aa}_{\aa,2k+1}=\aa;\\
  &b^{\aa}_{-\aa,2k}=-\aa,&&b^{\aa}_{-\aa,2k+1}=0.
\end{align}
We show the corresponding contours for $\aa=+$ in Fig.\,\ref{fig_BorelPlane}. Note that the contour choice depends on the label $\cc$ introduced in (\ref{eq_BorelIntegrand}). Now, we can sum contributions from different branch cuts and use (\ref{eq_PQconnection}) to get:
\begin{align}
  \G_{\text{on-shell}}^{(\cc)\aa}=&~\frac{\ii\aa e^{-\ii\aa\pi p_{12}/2}r_1r_2 e^{-\aa\cc\pi\wt\nu}}{2\sinh^2(\pi\wt\nu)(1-r_1^2)^{p_1/2+1}(1-r_2^2)^{p_2/2+1}}\n\\
  &\times\Big[\text{Q}^{p_1+2}_{-\ii\cc\wt\nu-1/2}(r_1^{-1})\text{Q}^{p_2+2}_{\ii\cc\wt\nu-1/2}(r_2^{-1})-\text{Q}^{p_1+2}_{-\ii\cc\wt\nu-1/2}(r_1^{-1})\text{Q}^{p_2+2}_{-\ii\cc\wt\nu-1/2}(r_2^{-1})\Big].
\end{align}
It is straightforward to compare the result with what we found in Sec.\,\ref{sec_signal}. Clearly, the first term above has the form of the local signal and the second the nonlocal signal. We can again use the unitary condition to get the ``unitarized'' on-shell contribution as in (\ref{eq_Gunitarized}) and also the full result in (\ref{eq_Gunitary}).

\section{Resurgence with a General Mass Spectrum}
\label{sec_general}

Up to this point, for simplicity, we have been focusing on the resurgence of UV correlators with single massive exchange. However, the idea of recovering UV correlators via resummation plus boundary conditions can be applied to more general situations as well. In this section, we explore more general situations beyond single exchange processes and consider two representative classes of generalizations. First, we show in Sec.\,\ref{sec_superPos} that the resummation can be performed for the single exchange of an arbitrary but finite discrete heavy mass spectrum. Second, we discuss in Sec.\,\ref{sec_generalTree} the recovery of UV correlators with arbitrary tree topologies, with an arbitrary number of massive exchanges and arbitrary couplings. Throughout this section, we will use the spectral resummation.

\subsection{Single exchange of a mass spectrum}
\label{sec_superPos}

As the first general case, we consider the single exchange diagram, still the same as the right panel of Fig.\;\ref{fig_resurgence}, but the internal line now represents the exchange of $M$ heavy particles with mass parameters $\wt\nu_\al$ ($\al=1,\cdots,M)$. 

As a simple extension of our single-exchange model in (\ref{eq_Lag}) we consider the UV model with the following Lagrangian:
\bge
  \ld' =-\sum_{\al=1}^M\FR{1}{2}\bigg[a^2(\pd_\mu\si_\al)^2+a^4m_\al^2\si_\al^2\bigg]-\FR{1}{2}\sum_{I=1}^2\bigg[a^2 (\pd_\mu\phi_I)^2+2a^4 \phi_I^2+\sum_{\al=1}^M\lam_{I,\al} a^{-p_I+2}\phi_I^2\si_\al\bigg],
\ede

where $m_\al^2=\wt\nu_\al^2+9/4$ are masses, and $\lam_{I,\al}$ are flavor dependent coupling strengths. In principle, we can be more general and make the time dependence of the couplings flavor-dependent but we do not bother to include this unnecessary complication.  

After integrating out all these heavy particles $\si_\al$, we get the following EFT Lagrangian:
\begin{align}
\label{eq_multiEFT}
  \ld_\text{EFT}'
  =-\FR{1}{2}\sum_{I=1}^2\Big[a^2 (\pd_\mu\phi_I)^2+2a^4 \phi_I^2 \Big]+\sum_{n=0}^\infty\Big[\mathcal{O}_{n}'+\mathcal{O}^{(1)}_{n}{}'+\mathcal{O}^{(2)}_{n}{}'\Big],
\end{align}
where we have three series of EFT operators: 
\begin{align}
   \mathcal{O}_n'=\left(\sum_{\alpha=1}^{M}\FR{\lam_{1,\alpha}\lam_{2,\alpha}}{4\wt\nu_\alpha^{2n+2}}\right)(-1)^na^4 (-\tau)^{p_1+2}\phi_1^2\bigg[\Big(\vartheta_\tau-\FR32\Big)^2-\tau^2\pd_i^2\bigg]^n\Big[(-\tau)^{p_2+2}\phi_2^2\Big],\\
  \mathcal{O}^{(1)}_n{}'=\left(\sum_{\alpha=1}^{M}\FR{\lam_{1,\alpha}^2}{8\wt\nu_\alpha^{2n+2}}\right)(-1)^na^4(-\tau)^{p_1+2}\phi_1^2\bigg[\Big(\vartheta_\tau-\FR32\Big)^2-\tau^2\pd_i^2\bigg]^n\Big[(-\tau)^{p_1+2}\phi_1^2\Big],\\
   \mathcal{O}^{(2)}_n{}'=\left(\sum_{\alpha=1}^{M}\FR{\lam_{2,\alpha}^2}{8\wt\nu_\alpha^{2n+2}}\right)(-1)^na^4(-\tau)^{p_2+2}\phi_2^2\bigg[\Big(\vartheta_\tau-\FR32\Big)^2-\tau^2\pd_i^2\bigg]^n\Big[(-\tau)^{p_2+2}\phi_2^2\Big].
\end{align}
Once again, only operators $\{\mathcal{O}_n'\}_{n=0}^{\infty}$ contribute to the correlator $\la \phi_1\phi_1\phi_2\phi_2\ra'$, and the corresponding contribution $\G_n'(r_1,r_2)$ from the $n$-th operator $\mathcal{O}_n'$ is:
\begin{align}
  \G_n'(r_1,r_2)=\sum_{\alpha=1}^{M}\lambda_{1,\alpha}\lambda_{2,\alpha}\G_n^{(\wt\nu_\al)}(r_1,r_2).
\end{align}
Here, we have kept the coupling coefficients, and $\G_n^{(\wt\nu)}(r_1,r_2)$ is the EFT correlator given in (\ref{eq_Gn}) with all $\wt\nu$ replaced by $\wt\nu_\al$. 

Although the equations look similar, there is a difference from the previous single-exchange case, where the EFT can be understood as a power expansion in $1/\wt\nu^2$. Here, the $n$-th order of the EFT expansion $\propto 1/\wt\nu^{2n+2}$ is not directly associated with the expansion over a single mass parameter. On the other hand, we know that the corresponding EFT expansion in flat spacetime can be understood as a gradient expansion, in powers of $p^2$, which is the eigenvalue of the quadratic Casimir operator of Poincaré group. Similarly, we can understand the current EFT with multiple flavors integrated out as an expansion in powers of spectral variable $\mu^2$ which is the eigenvalue of the quadratic Casimir operator of dS isometry group. Indeed, from (\ref{eq_rho_n}), we can immediately extract the spectral density of $\G_n'(r_1,r_2)$ (See also (\ref{eq_RepU})):
\begin{align}
    \rho_n'(\mu)=-\sum_{\al=1}^{M}\lambda_{1,\alpha}\lambda_{2,\alpha}\frac{\mu\sinh(\pi\mu)}{\pi\wt\nu_\al^2}\left(\frac{\mu}{\wt\nu_\al}\right)^{2n}=\frac{\mu\sinh(\pi\mu)}{\pi}\mathcal{A}_n(\mu^2).
\end{align}
Here, the normalized density $\mathcal{A}_n(p^2)$ is nothing but the $n$-th order scattering amplitude of EFT in flat spacetime after integrating out $M$ heavy particles. So we see explicitly that all terms in $\mathcal{A}_n(\mu^2)$ share the same power of $\mu^2$. Also, the resummation is straightforward:
\begin{align}
\label{eq_GprimeResummed}
  \G'(r_1,r_2)\sim \sum_{\aa=\pm} \ii\aa \int_{-\infty}^{+\infty}\di \mu\,\frac{\mu\sinh(\pi\mu)}{\pi}\mathcal{A}(\mu^2)\mathcal{U}^{p_1}_\aa(\mu;r_1)\mathcal{U}^{p_2}_\aa(\mu;r_2),
\end{align}
where
\begin{align}
\label{eq_QFTA}
  \mathcal{A}(\mu^2)=\sum_{n=0}^\infty\mathcal{A}_n(\mu^2)=\sum_{\alpha=1}^{M}\frac{\lambda_{1,\alpha}\lambda_{2,\alpha}}{\mu^2-\wt\nu_\al^2}
\end{align}
Then, from the UV amplitude $\mathcal{A}(s)$ in flat spacetime, we can construct the full UV correlator in dS by finishing the spectral integral. Moreover, we can recognize $M$ different UV degrees of freedom as $M$ poles of $\mathcal{A}(\mu^2)$. We can impose the boundary condition for each of the poles with an appropriate $\ii\epsilon$-prescription. The treatment parallels the previous single exchange case so we will not repeat ourselves. The upshot is that the resurgence of a mass spectrum can be manifestly realized once we know the spectral representation of EFT correlators and the pole structure of the resummed spectral density.

\paragraph{Padé resummation of truncated EFT series}
While the above extension to $M$ copies of heavy states is somewhat trivial, we can ask a more interesting and more practical question: Can we reconstruct the UV correlators approximately from a truncated EFT expansion? Since the spectral representation has translated correlators into flat-space scattering amplitudes as in (\ref{eq_GprimeResummed}), we can answer the question by importing known results for scattering amplitudes.

The approximate reconstruction of UV states was a well studied topic in scattering theory. In particular, it was shown recently in \cite{Calisto:2026pvv} that the UV spectrum can be reconstructed approximately and inductively from a finite number of Wilson coefficients in the flat-space EFT. The essential idea is simple: A Taylor polynomial of degree $N$ uniquely determines its Padé approximation of order $[N_1/N_2]$ satisfying $N_1+N_2=N$. In particular, for our example in (\ref{eq_QFTA}), $\mathcal{A}(\mu^2)$ happens to be a rational function of $\mu^2$. So, we can determine the UV mass spectrum precisely given that we know that there are $M$ states in the UV and that we know the Taylor coefficients of the (normalized) spectral density $\mathcal{A}(\mu^2)$ up to $O(\mu^{4M})$.

Specifically, suppose that the EFT expansion of the spectral density has the following form:
\begin{align}
\label{eq_Aseries}
  \mathcal{A}(\mu^2)=\sum_{k=0}^{\infty}a_k \mu^{2k}.
\end{align}
Here, for simplicity, we assume that $\mathcal{A}$ is regular at $\mu^2=0$. Also, suppose that we only know the first $2\ell$ terms of (\ref{eq_Aseries}). Then, we can construct the following matrices from these Wilson (or Taylor) coefficients $a_k$:
\begin{align}
  &\left[\mathbf{A}_0^{(\ell)}\right]_{ij}=a_{i+j},&&\left[\mathbf{A}_1^{(\ell)}\right]_{ij}=a_{i+j+1},&&(0\le i,j\le \ell-1).
\end{align}
Then, the Padé approximation of order $[(\ell-1)/\ell]$ can be constructed by solving for $\rho$ from the following equation:
\begin{align}
\label{eq_det}
  \text{Det}\left(\mathbf{A}_0^{(\ell)}-\rho\mathbf{A}_1^{(\ell)}\right)=0.
\end{align}
As an equation of $\ell$'s power in $\rho$, it generally produces $\ell$ distinct solutions 
$\rho=\rho_{1}^{(\ell)},\cdots,\rho_{\ell}^{(\ell)}$. Then, the Padé approximation is constructed as:
\begin{align}
  \mathcal{A}(\mu^2)\sim\frac{\sum_{k=0}^{\ell-1}b_k\mu^{2k}}{(\mu^2-\rho_{1}^{(\ell)})(\mu^2-\rho_{2}^{(\ell)})\cdots(\mu^2-\rho_{\ell}^{(\ell)})},
\end{align}
where
\begin{align}
  b_k=\sum_{m=0}^{k}\frac{a_{k-m}}{m!}\frac{\di^m}{\di (\mu^2)^m}\left[(\mu^2-\rho_{1}^{(\ell)})\cdots(\mu^2-\rho_{\ell}^{(\ell)})\right]\Big|_{\mu^2=0}.
\end{align}
So, in this way, we have approximately recovered $\ell$ UV states. As $\ell$ grows, the result will get closer to the original UV expression with the precise mass spectrum. Eventually, we get precisely $M$ heavy particles when $\ell=M$. When $\ell>M$, we will find that $\text{Det}\,\mathbf{A}_{0,1}^{(\ell)}=0$, indicating a finite mass spectrum in the UV. Finally, the above algorithm can be translated back to the correlator level through (\ref{eq_GprimeResummed}), and we expect it to yield a useful numerical method for cosmological correlators. We leave this very interesting topic for future work.

\subsection{General tree graphs}
\label{sec_generalTree}

In this subsection, we will not limit ourselves to the single-massive-exchange correlator in the UV theory, but consider the exchanging process with an arbitrary tree graph in the bulk. For definiteness, we will still consider one graph at a time, say a tree graph $\G_V$ with $I$ internal lines and $V=I+1$ bulk vertices. In general, there could be many graphs contributing to the same correlator, but we can get rid of this difficulty by introducing the model in a somewhat ad hoc way, similar to what we have done previously. So, we consider a model of $V$ flavors of conformal scalars $\phi_i$ ($i=1,\cdots, V$) and $I$ flavors of heavy scalars $\si_\al$ ($\al=1,\cdots, I$), together with the following interactions:
\begin{align}
  &\ld_{\text{int}}=-\FR12\sum_{i=1}^{V}\lambda_i a^{-p_i+2}\phi_{i}^2\prod_{\alpha\in \mathcal{I}_i}\si_\alpha.
\end{align}
where $\mathcal{I}_i$ is an index set of all internal lines attached to Vertex $i$. In this model, let us consider the $2V$-point correlator $\la \phi_1\phi_1\cdots\phi_V\phi_V\ra'$, and it is straightforward to observe that the only graphic contribution at the tree level is precisely $\wh\G$. 

The full analytic result for $\G$ has been worked out in terms of MFT integrals and their dressings in \cite{Liu:2024str}. It would be useful to have basic knowledge about this result for understanding the following discussions. A succinct summary of it can be found in App.\;A of \cite{Xianyu:2025lbk}. Here, we only introduce notations for kinematic data of $\G_V$. As in \cite{Liu:2024str}, we use the line energy $K_\al$ to denote the magnitude of momentum flowing in Line $\al$ and the vertex energy $E_i$ to denote the magnitude sum of all momenta flowing in the external conformal scalar modes attached to Vertex $i$. Then, a properly normalized dimensionless graph $\G_V$ is a function of $2I$ ratios $r_{(\al i)}\equiv K_\al/E_i$.

As before, we can integrate out all the heavy scalars to get an EFT. Of course, we can integrate out those states in different orders, but the correlators computed from the resulting EFT should be independent of this order choice. To see this point explicitly, let us think of the integrating-out operation as an inductive procedure: Suppose that we have integrated out all heavy particles except $\si_\al$, so that there are only two types of interactions left:
\begin{align}
  \ld_{\text{int}}=a^4\left(\mathcal{O}_L\si_\al+\si_\al\mathcal{O}_R\right).
\end{align}
Here $\mathcal{O}_L$ and $\mathcal{O}_R$ are low-energy local operators of two subgraphs (left and right) separated by Line $\alpha$ of $\wh\G$, and both of them are a tower of EFT interactions labeled by $\{n_1,\cdots,n_{\alpha-1}\}$ and $\{n_{\alpha+1},\cdots,n_I\}$. So the final EFT operators after integrating out $\si_\al$ are found to be:
\begin{align}
  \mathcal{O}_{n_1,\cdots,n_I}=\FR12 a^4 \frac{(-1)^{n_\al}}{\wt\nu_\al^{2n_\al+2}}\left(\mathcal{O}_L+\mathcal{O}_R\right)\bigg[\Big(\vartheta_\tau-\FR32\Big)^2-\tau^2\pd_i^2\bigg]^{n_\al}\left(\mathcal{O}_L+\mathcal{O}_R\right).
\end{align}
Inductively, we can expand $\mathcal{O}_L$ and $\mathcal{O}_R$ in the same form, until every subgraph is a single vertex corresponding to the UV operator $-\frac{\lambda_i}{2}(-\tau)^{p_i+2}\phi_i^2$. For example, the EFT operator of the four-site star can be constructed in this way to be:
\begin{align}
  \vcenter{\hbox{\includegraphics{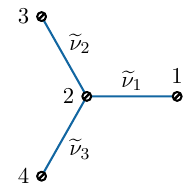}}}\sim&~ \frac{\lambda_1\lambda_2\lambda_3\lambda_4}{16}a^4\sum_{n_1,n_2,n_3=0}^{\infty}\prod_{\alpha=1}^{3}\frac{(-1)^{n_\al}}{\wt\nu_\al^{2n_\al+2}}(-\tau)^{p_1+2}\phi_1^2\bigg[\Big(\vartheta_\tau-\FR32\Big)^2-\tau^2\pd_i^2\bigg]^{n_1}\n\\
  &\times\Bigg\{(-\tau)^{p_2+2}\phi_2^2\bigg[\Big(\vartheta_\tau-\FR32\Big)^2-\tau^2\pd_i^2\bigg]^{n_2}\bigg[(-\tau)^{p_3+2}\phi_3^2\bigg]\n\\
  &\times\bigg[\Big(\vartheta_\tau-\FR32\Big)^2-\tau^2\pd_i^2\bigg]^{n_3}\bigg[(-\tau)^{p_4+2}\phi_4^2\bigg]\Bigg\}+\cdots.
\end{align}
Here $\cdots$ denotes operators that do not contribute to $\la\phi_1\phi_1\phi_2\phi_2\phi_3\phi_3\phi_4\phi_4\ra'$ at the tree level. Obviously, the form of these EFT operators varies as we change the order of the reduction, but they are equivalent through IBP as long as our theory is IR finite. More explicitly, we can work out the expression of the dimensionless correlator at fixed orders of $(n_1,n_2,n_3)$: 
\begin{align}
\label{eq_Gn1n2n3}
  \G_{n_1,n_2,n_3}=&\left(r_{(12)}r_{(23)}r_{(34)}\right)^3\prod_{\alpha=1}^{3}\frac{(-1)^{n_\al}}{\wt\nu_\al^{2n_\al+2}}\sum_{\aa=\pm}\ii\aa \int_{-\infty}^0\di z_1\,(-z_1)^{p_1}e^{\ii\aa z_1}\,\mathscr{D}_{K_1z_1/E_1}^{n_1}\Bigg\{\left(-\frac{E_2}{E_1}z_1\right)^{p_2+4}e^{\ii\aa \frac{E_2}{E_1}z_1}\n\\
  &\times\mathscr{D}_{K_2z_1/E_1}^{n_2}\Bigg[\left(-\frac{E_3}{E_1}z_1\right)^{p_3+4}e^{\ii\aa \frac{E_3}{E_1}z_1}\Bigg]\mathscr{D}_{K_3z_1/E_1}^{n_3}\Bigg[\left(-\frac{E_4}{E_1}z_1\right)^{p_4+4}e^{\ii\aa \frac{E_4}{E_1}z_1}\Bigg]\Bigg\}\n\\
  =&\prod_{\alpha=1}^{3}\frac{(-1)^{n_\al}}{\wt\nu_\al^{2n_\al+2}}\sum_{\aa=\pm}\ii\aa \int_{-\infty}^0\left[\prod_{i=1}^4\di z_i\,(-z_i)^{p_i}e^{\ii\aa z_i}\right](r_{(12)}z_2)^4(r_{(23)}z_3)^4(r_{(34)}z_4)^4\n\\
  &\times\left[\mathscr{D}_{r_{(11)}z_1}^{n_1}\delta(r_{(11)}z_1-r_{(12)}z_2)\right]\left[\mathscr{D}_{r_{(22)}z_2}^{n_2}\delta(r_{(22)}z_2-r_{(23)}z_3)\right]\left[\mathscr{D}_{r_{(32)}z_2}^{n_3}\delta(r_{(32)}z_2-r_{(34)}z_4)\right]\n\\
  =&\sum_{\aa=\pm}\ii\aa\int_{-\infty}^{\infty}\left[\prod_{\al=1}^3\di\mu_\al\,\frac{-\mu_\al\sinh(\pi\mu_\al)}{\pi\wt\nu_\al^2}\left(\frac{\mu_\al}{\wt\nu_\al}\right)^{2n_\al}\right]\n\\
  &\times\mathcal{U}_\aa^{p_1}(\mu_1;r_{(11)})\mathcal{U}_\aa^{p_2}(\mu_1,\mu_2,\mu_3;r_{(12)},r_{(22)},r_{(32)})\mathcal{U}_\aa^{p_3}(\mu_2;r_{(23)})\mathcal{U}_\aa^{p_4}(\mu_3;r_{(34)}).
\end{align}
Here, the differential operator $\mathcal{D}_{r z}$ is introduced in (\ref{eq_Dz}), and in the last line, we have used the orthonormal relation (\ref{eq_OrthNormS}) to express three $\delta$-functions and introduced the $N$-point function as the generalization of (\ref{eq_3pt}):
\begin{align}
  &\mathcal{U}_\aa^{p}(\mu_1,\cdots,\mu_N;r_1,\cdots,r_N)=\int_{-\infty}^0\di z (-z)^pe^{\ii\aa z}\prod_{\al=1}^{N}S_{\aa}(\mu_\al;r_\al z)\n\\
  =&\left(\frac{2}{\pi}\right)^{N/2}e^{-\ii\aa(p+N/2+1)\pi/2}\sum_{\cc_1,\cdots,\cc_N=\pm}\Gamma[-\ii\cc_1\mu_1,\cdots,-\ii\cc_N\mu_N,p+\ii(\cc\mu)_{1\cdots N}+3N/2+1]\n\\
  &\times\prod_{\al=1}^{N}\left(\frac{r_\al}{2}\right)^{\ii\cc_\al\mu_\al+3/2}\mathrm{F}_{\mathrm{C}}\left[\bgm \frac{p+\ii(\cc\mu)_{1\cdots N}+1}{2}+\frac{3N}{4},\frac{p+\ii(\cc\mu)_{1\cdots N}+2}{2}+\frac{3N}{4}\\1+\ii\cc_1\mu_1,\cdots,1+\ii\cc_N\mu_N\edm\Bigg|r_1^2,\cdots,r_N^2\right].
\end{align}
The definition of Lauricella's $\mathrm{F}_{\mathrm{C}}$ function can be found in App.\,\ref{app_2F1}.

From the final expression of (\ref{eq_Gn1n2n3}), we see that the EFT correlator at a given order is an integral over products of ``$N$-point functions'' weighted by expanded spectral density functions. This is still true in general situations: For an arbitrary tree graph, the EFT result at an given order $(n_1,\cdots,n_I)$ can be written as: 
\begin{align}
  \G_{n_1,\cdots,n_I}\Big(\{r_{(\al i)}\}\Big)=\sum_{\aa=\pm}\ii\aa\int_{-\infty}^{\infty}\left[\prod_{\al=1}^I\di\mu_\al\, \rho_{n_\al}(\wt\nu_\al;\mu_\al)\right]\prod_{i=1}^V\mathcal{U}_\aa^{p_i}\Big(\{\mu_\al\}_{\al\in\mathcal{I}_i};\{r_{(\al i)}\}_{\al\in\mathcal{I}_i}\Big),
\end{align}
where $\mathcal{I}_i$ denotes the set of lines connected to Vertex $i$, and $\rho_{n_\al}(\wt\nu_\al;\mu_\al)=-[\mu_\al\sinh(\pi\mu_\al)/(\pi\wt\nu_\al^2)]\times(\mu_\al/\wt\nu_\al)^{2n_\al}$. 
Then, the resummation of EFT graphs can be performed similarly:
\begin{align}
\label{eq_MFTsRep}
  \G\Big(\{r_{(\al i)}\}\Big)\sim&\sum_{n_1,\cdots,n_I=0}^{\infty}\G_{n_1,\cdots,n_I}\Big(\{r_{(\al i)}\}\Big)\n\\
  \sim&\sum_{\aa=\pm}\ii\aa\int_{-\infty}^{\infty}\left[\prod_{\al=1}^I\di\mu_\al\, \rho_\aa(\wt\nu_\al;\mu_\al)\right]\prod_{i=1}^V\mathcal{U}_\aa^{p_i}\Big(\{\mu_\al\}_{\al\in\mathcal{I}_i};\{r_{(\al i)}\}_{\al\in\mathcal{I}_i}\Big).
\end{align}
By requiring the standard boundary conditions, the spectral density $\rho_\aa(\wt\nu_\al;\mu_\al)$ is just what we found in (\ref{eq_rhoUV}) and (\ref{eq_beta}). This spectral representation of general massive trees was also worked out in \cite{Belrhali:2026act,Grafe:2026avi}. The spectral integral (\ref{eq_MFTsRep}) can be finished by properly closing the contours and picking up residues of enclosed poles. We spell out the details in App.\;\ref{app_general_tree} and show here the result for the background, which is the part of the whole answer suppressed by powers of internal masses: 
\begin{align}
\label{eq_generalBG}
  \G_{\text{BG}}=&~2^{3V-2}\cos(\pi p_{1\cdots V}/2)\sum_{\ell_2,\cdots,\ell_V=0}^{\infty}\Gamma(p_{1\cdots V}+\ell_{2\cdots V}+4V-3)\n\\
  &\times \mathrm{F}_{\mathrm{C}}\left[\bgm \frac{\ell_{2\cdots V}+p_{1\cdots V}+4V-3}{2},\frac{\ell_{2\cdots V}+p_{1\cdots V}+4V-2}{2}\\\left\{\ell_j+p_j+q_j+7/2\right\}_{j\in\mathcal{I}_1}\edm\Bigg|\{r_{(j 1)}^2\}_{j\in\mathcal{I}_1}\right]\n\\
  &\times\prod_{i=2}^{V}\frac{(-1)^{\ell_i}}{\ell_i!\left[(\ell_i+p_i+q_i+5/2)^2+\wt\nu_i^2\right]}\left(\frac{K_i}{2E_1}\right)^3\left(\frac{E_i}{E_1}\right)^{\ell_i+p_i+1}\n\\
  &\times\mathrm{F}_{\mathrm{C}}\left[\bgm -\fr{\ell_i}{2},-\fr{\ell_i-1}{2}\\-\ell_i-p_i-q_i-3/2,\left\{\ell_j+p_j+q_j+7/2\right\}_{j\in\mathcal{I}_i/\{i\}}\edm\Bigg|\{r_{(j i)}^2\}_{j\in\mathcal{I}_i}\right].
\end{align}
When writing this result, we have chosen Vertex 1 to carry the largest vertex energy with $E_1>\sum_{i=2}^V E_i$, and as such, the graph $\G$ acquires a natural family-tree structure, with Vertex 1 being the root and all lines flowing outwards from the root. In this way, the family-tree structure is encoded in the following family parameter $q_i$: (See also \cite{Liu:2024str} for more detailed discussions.)
\begin{align}
  q_i\equiv \wt\ell_i+\wt p_i+4N_i,
\end{align}
where $\wt\ell_i$ means the sum of all $\ell_j$'s over the descendant vertices of Vertex $i$, $\wt p_i$ is defined
similarly, and $N_i$ is the number of descendant vertices of Vertex $i$. The similar result was also obtained in \cite{Grafe:2026avi}. Also, we prove in App.\,\ref{app_CIS} that the above expression is equivalent to the MFT for the completely inhomogeneous solution found in \cite{Liu:2024str}.

\section{Discussions and Outlooks}
\label{sec_conclusion}
The possibility of extracting nonperturbative information from a perturbative EFT Lagrangian is a problem of general interest in quantum field theory. In this work, we studied a particular incarnation of this problem in the context of cosmological correlators. In particular, we worked with a typical example of 4-point correlator with single exchange of a heavy scalar of mass $m$, and then considered the exchange of a general heavy mass spectrum as well as arbitrary exchanging correlators at the tree level. We showed that, after integrating out the heavy scalar, the low-energy EFT has sufficient information to recover the smooth background of the correlator. This is achieved by a proper resummation of the divergent EFT series which we performed in three different ways. After the resummation, we found new poles/cuts in the spectral density and the Borel transform, which give rise to the exponentially suppressed contributions to the correlator and correspond to the nonperturbative effect of cosmological particle production and are part of the CC signals. However, we also showed that the precise coefficients of these signals cannot be determined by the EFT alone, since they depend on the boundary conditions that characterize the heavy propagating degrees which are missing in the low-energy EFT. The additional information can be included by requiring the final-state unitarity and the BD initial state for the heavy particle. With these boundary conditions imposed, we finally recover the complete correlator in the UV theory.

Our exercise in this work can be viewed in combination with the dispersive bootstrap in \cite{Liu:2024xyi}. Altogether, they showed a bidirectional relation between the CC signal and background: We can either start from low-energy EFT to recover the CC background through resummation and also the signal by imposing appropriate boundary conditions, or start from the signal to dispersively bootstrap the full background modulo a finite number of counterterms. This one-to-one mapping provides an explicit realization of resurgence, showing that cosmological correlators can be a natural playground to study resurgent properties of cosmological EFTs. Clearly, we have only studied the simplest case with tree-level exchanges. It would be interesting to see if the resurgent relations uncovered here can be extended to more complicated situations such as loop exchanges. The spectral functions in these cases are certainly more complicated but may also possess new interesting structures not seen here. We leave this interesting topic for future work. 

At the same time, it is also important to note that the information stored in either the background or the signal is not completely lossless. In particular, for the example of single massive exchange, the low-energy EFT can only recover the $(\pm\pm)$ branches of the SK time integral of the UV correlator, which by themselves are not unitary from the viewpoint of summing over final states, a phenomenon already noticed in the literature \cite{Green:2024cmx}. It is tempting to try to include the information of $(\pm\mp)$ branches in the low-energy EFT by adding boundary terms to the Lagrangian which mixes field variables in the $+$ and $-$ branches. While this is expected from general considerations about EFTs in SK formalism \cite{Salcedo:2024smn}, the new boundary terms themselves would be nonlocal and suppressed by exponentially small coefficients in the large mass limit for cosmological correlators with massive exchanges, obscuring the structure of derivative expansion in a local EFT. Therefore, we choose to provide this information through an explicit requirement of final-state unitarity in the UV theory instead of adding boundary terms to low-energy EFT. Nevertheless, it could be interesting to try the alternative route, searching for an organizing principle for boundary terms in the low-energy EFT that allows us to systematically include nonperturbative boundary data directly in the EFT. We also leave this topic for future study.

\paragraph{Note added} During the completion of this work, we became aware of a forthcoming work by Daniele Dorigoni, Laura Engelbrecht, Nadine Nussbaumer, and Guilherme L.\ Pimentel, which independently studies similar questions.

\paragraph{Acknowledgments} We thank Laura Engelbrecht, Carlos Duaso Pueyo, Kamran Salehi Vaziri, Denis Werth, and Hongyu Zhang for useful discussions. This work is supported by NSFC under Grants No.\ 12275146 and No.\ 12247103, the National Key R\&D Program of China (2021YFC2203100), and the Dushi Program of Tsinghua University. 

\newpage
\begin{appendix}

\section{Recovery of the Background for General Trees}
\label{app_general_tree}

In this appendix, we spell out the details of finishing the spectral integral for a general tree graph in (\ref{eq_MFTsRep}) to obtain the full analytic background of the graph in (\ref{eq_generalBG}). For definiteness, we work in the region with $E_1>\sum_{i=2}^{V}E_i$. Then, Vertex 1, as the maximal energy site, naturally generates a family-tree structure to the tree graph with itself being the root. Then, we use the same conventions as in \cite{Liu:2024str} and require that an arbitrary Line $j$ always points from the mother vertex $j'$ to the daughter vertex $j$. Then, we explicitly finish spectral integrals in a specific order. That is, we always first finish integrals of $\mu_\al$ that $\al$ is in the index set of the earlier generation, and in this way we will show that $E_1>\sum_{i=2}^{V}E_i$ allows us to close the integration contour in the correct direction. 

Below, we first introduce a kinematic inequality that will be useful for following discussions. Then, we discuss the two cases we face in the whole computation, and it is easy to extract the general expression in an inductive manner.\footnote{It was also shown in \cite{Grafe:2026avi} that picking up  appropriate poles of the spectral integrand gives the correct answer in (\ref{eq_generalBG}).} 

\begin{figure}[t]
\centering
\includegraphics[width=0.6\textwidth]{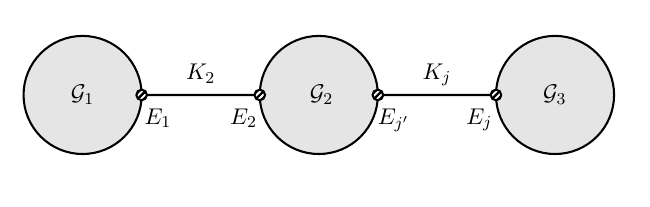}
\caption{An illustration for kinematic variables of $\G$.}
\label{fig_ineq}
\end{figure}

\paragraph{A kinematic inequality} 
Let us consider an arbitrary tree graph with line energies $K_2,\cdots,K_N$ and vertex energies $E_1,\cdots,E_V$. For physically reachable configurations, these kinematic variables satisfy the following sets of inequalities as long as $E_1>\sum_{i=2}^{V}E_i$:
\begin{align}
\label{eq_ineq}
  \ln\left(\frac{K_2}{K_j}\right)+\xi\left(\frac{K_2}{E_1+\sum_{i\in\mathcal{I}_1/\{2\}}\dd_i K_i}\right)>\xi\left(\frac{K_j}{E_j+\sum_{i\in\mathcal{I}_j/\{j\}}K_i}\right),
\end{align}
where $\dd_i=\pm$, $\xi(z)\equiv \text{arccosh}\,(1/z)$, and $\mathcal{I}_j$ is, as introduced in the main text, the set of lines connected to Vertex $j$. As shown in Fig.\,\ref{fig_ineq}, we have supposed that Vertex $2$ is the daughter of Vertex $1$, and Vertex $j$ can be any descendant of Vertex $2$. Then, by removing Line $2$ and Line $j$, the whole graph is separated into three disconnected parts $\G_1,\G_2,\G_3$. Correspondingly, we have the following three triangle inequalities due to momentum conservation:
\begin{align}
  \sum_{i\in\G_1}E_i&\ge E_1+\sum_{i\in\mathcal{I}_1/\{2\}}K_i;\\
  \sum_{i\in\G_2}E_i&\ge |K_2-K_j|;\\
  \sum_{i\in\G_3}E_i&\ge E_j+\sum_{i\in\mathcal{I}_j/\{j\}}K_i.
\end{align}
Also, our assumption of $E_1\geq \sum_{i=2}^V E_i$ leads to:
\begin{align}
  E_1>\sum_{i=2}^{V}E_i=\sum_{i\in\G_1}E_i-E_1+\sum_{i\in\G_2}E_i+\sum_{i\in\G_3}E_i.
\end{align}
Then, combining these 4 inequalities, we can easily get:
\begin{align}
  E_1+\sum_{i\in\mathcal{I}_1/\{2\}}\dd_i K_i-E_j-\sum_{i\in\mathcal{I}_j/\{j\}}K_i>|K_2-K_j|\ge 0,
\end{align}
which further leads to:
\begin{align}
\label{eq_ineq0}
  \left(E_1+\sum_{i\in\mathcal{I}_1/\{2\}}\dd_i K_i\right)^2-K_2^2>\left(E_j+\sum_{i\in\mathcal{I}_j/\{j\}}K_i\right)^2-K_j^2.
\end{align}
From this inequality, it is easy to check that the following inequality also holds:
\begin{align}
\label{eq_inequlityLog}
  \ln\left(\frac{\sqrt{(E_1+\sum_{i\in\mathcal{I}_1/\{2\}}\dd_i K_i)^2-K_2^2}+E_1+\sum_{i\in\mathcal{I}_1/\{2\}}\dd_i K_i}{\sqrt{(E_j+\sum_{i\in\mathcal{I}_j/\{j\}}K_i)^2-K_j^2}+E_j+\sum_{i\in\mathcal{I}_j/\{j\}}K_i}\right)>0.
\end{align}
On the other hand, by using the identity for $\xi(r)$,
\begin{align}
  \xi(r)=\text{arccosh}\,\left(\frac{1}{r}\right)=\ln\left(\frac{\sqrt{1-r^2}+1}{r}\right)\quad (\Re(r)>0),
\end{align}
it is also easy to check that (\ref{eq_inequlityLog}) is equivalent to the original inequality in
(\ref{eq_ineq}), and therefore the original inequality is proved. 

The inequality (\ref{eq_ineq}) is useful for us to decide how to close the spectral contour in the following computation. This is naturally separated into two cases, which we discuss in turn.

\begin{figure}[t]
\centering
\includegraphics[width=0.4\textwidth]{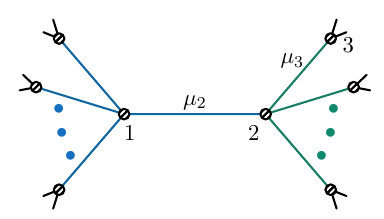}
\caption{The general tree structure of the UV correlator. Here, Vertex 1 (the root) carries the maximum vertex energy. The spectral integrals of blue lines around the root correspond to Case 1, and that of green lines Case 2. }
\label{fig_MFT}
\end{figure}

\paragraph{Case 1: integrals around the root}Without loss of generality, we first consider Line $2$, which points from the root Vertex $1$ to the daughter Vertex $2$, as shown in Fig.\,\ref{fig_MFT}. The relevant part of the spectral representation for the $\mu_2$-integral is:
\begin{align}
  \G=\cdots\times\int_{-\infty}^{\infty}\di\mu_2\,\rho_{\aa}(\wt\nu_2;\mu_2)\mathcal{U}_{\aa}^{p_1}\left(\{\mu_i\}_{i\in\mathcal{I}_1};\{r_{(i1)}\}_{i\in\mathcal{I}_1}\right)\mathcal{U}_{\aa}^{p_2}\left(\{\mu_i\}_{i\in\mathcal{I}_2};\{r_{(i2)}\}_{i\in\mathcal{I}_2}\right).
\end{align}
Similar to Sec.\,\ref{sec_compute_contact} and Sec.\,\ref{sec_signal}, for $j\in\mathcal{I}_k$ the large-$\mu_j$ behavior of the vertex function is:
\begin{align}
\label{eq_limitU}
  &\mathcal{U}_{\aa}^{p_k}\left(\{\mu_i\}_{i\in\mathcal{I}_k};\{r_{(ik)}\}_{i\in\mathcal{I}_k}\right)\n\\
  &\sim e^{-\pi\mu_j}\sum_{\cc_j=\pm}\ii\cc_j(\ii\cc_j\mu_j)^{p_k+M_k+1/2}e^{-\ii\cc_j\mu_j\xi\left(\frac{K_j}{E_k+\sum_{i\in\mathcal{I}_k/\{j\}}K_i}\right)},\qquad |\mu_j|\to\infty.
\end{align}
Here $M_k$ is the total number of internal lines attached to Vertex $k$. Such expression is observed from the asymptotic expansion of Lauricella functions $\mathrm{F}_{\mathrm{C}}$ in (\ref{eq_FCasymp}). For Line $2$, we denote its indices $\cc=\pm$ in $\mathcal{U}_{\aa}^{p_1}$ and $\mathcal{U}_{\aa}^{p_2}$ by $\cc_2$ and $\bar\cc_2$, respectively. Then, according to the inequality (\ref{eq_ineq}), the integration contour can be closed in the $\Im(\cc\mu_2)<0$ half-plane for both $\cc_2=\bar\cc_2=\cc$ and $\cc_2=-\bar\cc_2=\cc$. The corresponding poles are collected in Table \ref{tab_mu2pole}, where $\ell_2=0,1,\cdots$.
\begin{table}[t]
\centering
\caption{All relevant poles in the $\Im(\cc\mu_2)<0$ half-plane}
\label{tab_mu2pole}
\vspace{4mm}
\begin{tabular}{|c|cc|}
\hline
Poles                          & \multicolumn{1}{c|}{$\cc_2=\ob\cc_2=\cc$}  & $\cc_2=-\ob\cc_2=\cc$                  \\ \hline
\multirow{2}{*}{Off-shell}     & \multicolumn{2}{c|}{$\mu_2=-\ii\cc(\ell_2+1)$}                 \\ \cline{2-3}
                               & \multicolumn{1}{c|}{-}                  & $\mu_2=\cc\sum_{i\in\mathcal{I}_2/\{2\}}\cc_i\mu_i-\ii\cc(\ell_2+p_2+3M_2/2+1)$ \\ \hline
\multicolumn{1}{|l|}{On-shell} & \multicolumn{2}{c|}{$\mu_2=\pm\wt\nu_2-\ii\cc\epsilon$}                        \\ \hline
\end{tabular}
\end{table}

We will only consider off-shell contributions here to capture the full analytic background of the UV correlator. Using the limiting form of Lauricella function $\mathrm{F}_{\mathrm{C}}$ in (\ref{eq_FCn}), we again find that the contributions of $\mu_2=-\ii\cc(\ell_2+1)$ from the two cases exactly cancel. Then, the complete result of the $\mu_2$-integral is found to be:
\begin{align}
\label{eq_mu2Int}
  &\int_{-\infty}^{\infty}\di\mu_2\,\rho_{\aa}(\wt\nu_2;\mu_2)\mathcal{U}_{\aa}^{p_1}\left(\{\mu_i\}_{i\in\mathcal{I}_1};\{r_{(i1)}\}_{i\in\mathcal{I}_1}\right)\mathcal{U}_{\aa}^{p_2}\left(\{\mu_i\}_{i\in\mathcal{I}_2};\{r_{(i2)}\}_{i\in\mathcal{I}_2}\right)\n\\
  =&-\left(\frac{2}{\pi}\right)^{\frac{M_{12}-2}{2}}e^{-\ii\aa(p_{12}+M_{12}/2+2)\pi/2}r_{(21)}^3\n\\
  &\times\sum_{\{\cc\}=\pm}\left[\prod_{i\in\mathcal{I}_1/\{2\}}\Gamma\left[-\ii\cc_i\mu_i\right]\left(\frac{r_{(i1)}}{2}\right)^{\ii\cc_i\mu_i+3/2}\right]\left[\prod_{i\in\mathcal{I}_2/\{2\}}\Gamma\left[-\ii\cc_i\mu_i\right]\left(\frac{r_{(i2)}}{2}\right)^{\ii\cc_i\mu_i+3/2}\right]\n\\
  &\times\sum_{\ell_2=0}^{\infty}\frac{(-1)^{\ell_2}}{\ell_2!}\frac{\Gamma\left(\ell_2+p_{12}+\ii\sum_{i\in\mathcal{I}_{12}/\{2\}}\cc_i\mu_i+3M_{12}/2+2\right)}{\left[\ell_2+p_2+\ii\sum_{i\in\mathcal{I}_2/\{2\}}\cc_i\mu_i+3M_2/2+1\right]^2+\wt\nu_2^2}\left(\frac{r_{(21)}}{r_{(22)}}\right)^{\ell_2+p_2+\ii\sum_{i\in\mathcal{I}_2/\{2\}}\cc_i\mu_i+(3M_2-1)/2}\n\\
  &\times\mathrm{F}_{\mathrm{C}}\left[\bgm \frac{\ell_2+p_{12}+\ii\sum_{i\in\mathcal{I}_{12}/\{2\}}\cc_i\mu_i+1}{2}+\frac{3M_{12}+2}{4},\frac{\ell_2+p_{12}+\ii\sum_{i\in\mathcal{I}_{12}/\{2\}}\cc_i\mu_i+2}{2}+\frac{3M_{12}+2}{4}\\\ell_2+p_2+\ii\sum_{i\in\mathcal{I}_2/\{2\}}\cc_i\mu_i+\frac{3M_2+4}{2},\left\{1+\ii\cc_i\mu_i\right\}_{i\in\mathcal{I}_1/\{2\}}\edm\Bigg|r_{(21)}^2,\{r_{(i1)}^2\}_{i\in\mathcal{I}_1/\{2\}}\right]\n\\
  &\times\mathrm{F}_{\mathrm{C}}\left[\bgm -\fr{\ell_2}{2},-\fr{\ell_2-1}{2}\\-\ell_2-p_2-\ii\sum_{i\in\mathcal{I}_2/\{2\}}\cc_i\mu_i-\frac{3M_2}{2},\left\{1+\ii\cc_i\mu_i\right\}_{i\in\mathcal{I}_2/\{2\}}\edm\Bigg|\{r_{(i2)}^2\}_{i\in\mathcal{I}_2}\right]+\text{on-shell contributions}.
\end{align}
Here, we define $\mathcal{I}_{12}\equiv\mathcal{I}_1\cup\mathcal{I}_2$. The last Lauricella function, coming from the vertex function $\mathcal{U}_{\aa}^{p_2}$, is terminating because of the two upper parameters $-\ell_2/2$ and $-(\ell_2-1)/2$, while the first Lauricella function from the vertex function of the root remains non-terminating. The same discussion applies to every line attached to Vertex $1$, and thus all spectral integrals around the root receive contributions from the following poles with $\cc_{k}=-\ob\cc_{k}=\cc$:
\begin{align}
\label{eq_poles}
  &\mu_k=\cc\sum_{i\in\mathcal{I}_k/\{k\}}\cc_i\mu_i-\ii\cc(\ell_k+p_k+3M_k/2+1),&&\ell_k=0,1,\cdots.
\end{align}
Here, Vertex $k$ can be any daughter of Vertex $1$. After taking the residues of these poles, the Lauricella functions associated with these daughter vertices are all terminating polynomials.

\paragraph{Case 2: integrals around descendants}Next, we consider Vertex $3$, which is taken to be a daughter of Vertex $2$ as shown in Fig.\,\ref{fig_MFT}. After finishing the $\mu_2$-integral above, the Lauricella function associated with Vertex $2$ is already a terminating polynomial. Consequently, it can produce only algebraic powers of $\mu_3$ and does not contribute any new exponential dependence as $|\mu_3|\to\infty$, and the only two factors that determine the exponential behavior of the $\mu_3$-integrand are therefore the unique infinite Lauricella function inherited from Vertex $1$ and the unintegrated vertex function $\mathcal{U}_{\aa}^{p_3}$ of Vertex $3$. Meanwhile, the momentum-ratio factor carried by the $\mu_3$-dependence is $\left(K_2/K_3\right)^{-\ii\cc_3\mu_3}$. So from (\ref{eq_FCasymp}), the $\mu_3$-integrand is dominated by:
\begin{align}
  e^{-\ii\mu_3\left[\cc_3\ln\left(\frac{K_2}{K_3}\right)+\cc_3\xi\left(\frac{K_2}{E_1+\sum_{i\in\mathcal{I}_1/\{2\}}\dd_iK_i}\right)+\ob\cc_3\xi\left(\frac{K_3}{E_3+\sum_{i\in\mathcal{I}_3/\{3\}}K_i}\right)\right]}.
\end{align}
According to the inequality (\ref{eq_ineq}) with $k=3$, we should again pick all poles in the $\Im(\cc\mu_3)<0$ half-plane for both $\cc_3=\ob\cc_3=\cc$ and $\cc_3=-\ob\cc_3=\cc$, and the pole structure is the same as in Case 1. The first kind of off-shell poles $\mu_3=-\ii\cc(\ell_3+1)$ contributes nothing due to (\ref{eq_FCn}), while poles of the second kind are located at:
\begin{align}
  &\mu_3=\cc\sum_{i\in\mathcal{I}_3/\{3\}}\cc_i\mu_i-\ii\cc(\ell_3+p_3+3M_3/2+1)&&\ell_3=0,1,\cdots.
\end{align}
After picking up these poles, the vertex function $\mathcal{U}_{\aa}^{p_3}$ becomes:
\begin{align}
  \mathrm{F}_{\mathrm{C}}\left[\bgm -\fr{\ell_3}{2},-\fr{\ell_3-1}{2}\\-\ell_3-p_3-\ii\sum_{i\in\mathcal{I}_3/\{3\}}\cc_i\mu_i-\frac{3M_3}{2},\left\{1+\ii\cc_i\mu_i\right\}_{i\in\mathcal{I}_3/\{3\}}\edm\Bigg|\{r_{(i3)}^2\}_{i\in\mathcal{I}_3}\right],
\end{align}
which is again a terminating polynomial.

The same structure then applies recursively to an arbitrary descendant Vertex $j$ of Vertex $2$: all Lauricella functions associated with the non-root vertices whose spectral integrals have already been performed are terminating polynomials, and hence do not modify the exponential dependence as $|\mu_j|\to\infty$. Meanwhile, the momentum-ratio factors accumulated along the unique path from Line $2$ to Line $j$ telescope, so all intermediate line energies cancel and leave $\left(K_2/K_j\right)^{-\ii\cc_j\mu_j}$.
Therefore the $\mu_j$-integrand is dominated by:
\begin{align}
  e^{-\ii\mu_j\left[\cc_j\ln\left(\frac{K_2}{K_j}\right)+\cc_j\xi\left(\frac{K_2}{E_1+\sum_{i\in\mathcal{I}_1/\{2\}}\dd_iK_i}\right)+\ob\cc_j\xi\left(\frac{K_j}{E_j+\sum_{i\in\mathcal{I}_j/\{j\}}K_i}\right)\right]}.
\end{align}
Similarly using the inequality (\ref{eq_ineq}), we close the contour in the $\Im(\cc\mu_j)<0$ half-plane. The poles $\mu_j=-\ii\cc(\ell_j+1)$ give no contribution due to (\ref{eq_FCn}), while the remaining poles are:
\begin{align}
  &\mu_j=\cc\sum_{i\in\mathcal{I}_j/\{j\}}\cc_i\mu_i-\ii\cc(\ell_j+p_j+3M_j/2+1),&&\ell_j=0,1,\cdots.
\end{align}
After taking these residues, the Lauricella function associated with Vertex $j$ also becomes a terminating polynomial, and the same procedure can be continued layer by layer. Therefore the poles are always those in (\ref{eq_poles}), where Vertex $j$ can now be any descendant of the root. In other words, now we know locations of poles for all spectral parameters. We only need to sum up all residues of off-shell poles of the original integrand in (\ref{eq_MFTsRep}), and finally get the result in (\ref{eq_generalBG}).

\section{Algebraic Manipulation of Series Expressions}
In this appendix, we collect explicit derivations of various identities for series that appear in the main text. In App.\,\ref{app_G0}, we prove the identity (\ref{eq_G0}), which provides two equivalent expressions for $\G_0$. Similarly, we prove the equivalence between the two expressions, (\ref{eq_BG1}) and (\ref{eq_GBG}), of the UV background $\G_{\text{BG}}$ in App.\,\ref{app_BG}. Since $\G_0$ is the $0$-th order of $\G_{\text{BG}}$ in the large mass expansion, the treatments are similar. As a further application of the same method, we will show in App.\,\ref{app_CIS} that the expressions for the UV background of general UV correlators found in (\ref{eq_generalBG}) are  equivalent to the CIS series in \cite{Liu:2024str}.\footnote{Such identities are dubbed ``magical identities'' in \cite{Grafe:2026avi}.}

\subsection{Equivalent expressions of $\G_0$}
\label{app_G0}
Here, we prove the equivalence between two expressions of the EFT correlator $\G_0$ in (\ref{eq_G0}). On the second line of (\ref{eq_G0}), the expression is a triple hypergeometric series with summations over $\ell=0,1,\cdots$ for $(r_1/r_2)^{\ell}$, $m_1=0,1,\cdots$ for $r_1^{2m_1}$, and $m_2=0,1,\cdots,\left\lfloor \ell/2\right\rfloor$ for $r_2^{2m_2}$. (See (\ref{eq_2F1}) for the series definition of ${}_2\text{F}_1$ function.) We can change the summation indices from $(\ell,m_1,m_2)$ to $(\ell'=\ell-2m_2,m=m_{12},m_2)$, which is a bijection for the summand at each order. Then the series can be rewritten as:
\begin{align}
    \G_0(r_1,r_2)=&\sum_{\ell',m=0}^{\infty}\frac{2(-1)^{\ell'}\cos\left(\pi p_{12}/2\right)\Gamma(\ell'+2m+p_{12}+5)}{4^m\wt\nu^2\ell'!}\left(\frac{r_1}{r_2}\right)^{\ell'+p_2+1}r_1^{2m+3}\n\\
    &\times\sum_{m_2=0}^{m}\frac{(-1)^{m_2}(\ell'+2m_2+p_2+5/2)}{(m-m_2)!m_2!(p_2+\ell'+m_2+5/2)_{m+1}}.
\end{align}
Next, we prove that the sum over $m_2$ on the second line vanishes as long as $m\neq 0$: 
\begin{align}
    c_{\ell'm}\equiv\sum_{m_2=0}^{m}\frac{(-1)^{m_2}(p_2+\ell'+2m_2+5/2)}{(m-m_2)!m_2!(p_2+\ell'+m_2+5/2)_{m+1}}=\delta_{m0}.
\end{align}
The key observation is that the Pochhammer factors can be rewritten as the product of $m$ differences of squares:
\begin{align}
\label{eq_sqDiff}
    &m_2!(m-m_2)!\times(p_2+\ell'+m_2+5/2)_{m+1}\n\\
    =&\left[(-1)^{m_2}\prod_{k=0}^{m_2-1}\prod_{k=m_2+1}^{m}(k-m_2)\right]\times\left[(p_2+\ell'+2m_2+5/2)\prod_{k=0}^{m_2-1}\prod_{k=m_2+1}^{m}(p_2+\ell'+m_2+k+5/2)\right]\n\\
    =&~(-1)^{m_2}4^{-m}(p_2+\ell'+2m_2+5/2)\prod_{k=0}^{m_2-1}\prod_{k=m_2+1}^{m}\left[(p_2+\ell'+2k+5/2)^2-(p_2+\ell'+2m_2+5/2)^2\right].
\end{align}
So, the summation can be rewritten as:
\begin{align}
   c_{\ell'm}=\sum_{m_2=0}^{m}\frac{4^{m}}{\prod_{k=0}^{m_2-1}\prod_{k=m_2+1}^{m}\left[(p_2+\ell'+2k+5/2)^2-(p_2+\ell'+2m_2+5/2)^2\right]}.
\end{align}
Such expression can be thought of as the coefficient of $t^m$ in the generating polynomial $g(t)$:
\begin{align}
    g(t)=(-4)^{m}\sum_{m_2=0}^{m} \prod_{k=0}^{m_2-1}\prod_{k=m_2+1}^{m}\frac{t-(p_2+\ell'+2k+5/2)^2}{\left[(p_2+\ell'+2m_2+5/2)^2-(p_2+\ell'+2k+5/2)^2\right]}.
\end{align}
However, as a polynomial of degree $m$, it is obvious that:
\begin{align}
    &g[(p_2+\ell'+2k+5/2)^2]=(-4)^{m},&& k=0,1,\cdots,m.
\end{align}
Therefore, we have $g(t)\equiv (-4)^m$, where the coefficient of $t^m$ is $0$. So the triple summation above can be further simplified and we can get the familiar form:
\begin{align}
    \G_0(r_1,r_2)=&~\frac{2}{\wt\nu^2}\cos(\pi p_{12}/2)\sum_{\ell'=0}^{\infty}\frac{(-1)^{\ell'}\Gamma(\ell'+p_{12}+5)}{\ell'!}r_1^3\left(\frac{r_1}{r_2}\right)^{\ell'+p_2+1}\n\\
    =&~\frac{2}{\wt\nu^2}\cos(\pi p_{12}/2)\Gamma(p_{12}+5)\frac{r_1^{p_2+4}r_2^{p_1+4}}{r_{12}^{p_{12}+5}}.
\end{align} 
This is exactly the identity in (\ref{eq_G0}).

\subsection{Equivalent expressions of $\G_{\text{BG}}$}
\label{app_BG}
We now prove the equivalence between (\ref{eq_BG1}) and (\ref{eq_GBG}) in a similar fashion to that in the previous subsection. Starting from (\ref{eq_BG1}), we first explicitly expand two hypergeometric series and use the identity $2^{2m}(a)_m(a+1/2)_m=(2a)_{2m}$ to simplify the triple series as:
\begin{align}
    \G_{\text{BG}}=\sum_{\ell,m_1=0}^{\infty}\sum_{m_2=0}^{\left\lfloor \ell/2\right\rfloor}&\frac{(-1)^{\ell+m_2}(\ell+p_2+5/2)\cos(\pi p_{12}/2)\Gamma(p_{12}+\ell+2m_1+5)}{2^{2m_{12}-1}m_1!m_2!(\ell-2m_2)![(p_2+\ell+5/2)^2+\wt\nu^2](p_2+\ell-m_2+5/2)_{m_{12}+1}}\n\\
    &\times r_1^{\ell+2m_1+p_2+4}r_2^{2m_2-\ell-p_2-1}.
\end{align}
Now, similar to App.\,\ref{app_G0}, we define two new summation indices $m=m_{12},\ell'=\ell-2m_2$, and rewrite the triple summation as:
\begin{align}
    \sum_{\ell,m_1=0}^{\infty}\sum_{m_2=0}^{\left\lfloor \ell/2\right\rfloor}\cdots\times r_1^{\ell+2m_1+p_2+4}r_2^{2m_2-\ell-p_2-1}=\sum_{\ell',m=0}^{\infty}\sum_{m_2=0}^{m}\cdots\times r_1^{\ell'+2m+p_2+4}r_2^{-\ell'-p_2-1}.
\end{align}
So, now, the finite sum over $m_2$ becomes independent of $r_1,r_2$:
\begin{align}
    c_{\ell'm}=\sum_{m_2=0}^{m}\frac{(-1)^{m_2}(p_2+\ell'+2m_2+5/2)}{m_2!(m-m_2)!(p_2+\ell'+m_2+5/2)_{m+1}[(p_2+\ell'+2m_2+5/2)^2+\wt\nu^2]}.
\end{align}
Again, we use the identity (\ref{eq_sqDiff}), and then the $m_2$-sum becomes:
\begin{align}
\label{eq_clm}
    c_{\ell'm}=&\sum_{m_2=0}^{m}\frac{4^{m}\prod_{k=0}^{m_2-1}\prod_{k=m_2+1}^{m}\left[(p_2+\ell'+2k+5/2)^2-(p_2+\ell'+2m_2+5/2)^2\right]^{-1}}{(p_2+\ell'+2m_2+5/2)^2+\wt\nu^2}\n\\
    =&4^{m}\prod_{k=0}^{m}\frac{1}{(p_2+\ell'+2k+5/2)^2+\wt\nu^2}=\frac{1}{4\big(\fr{p_2+\ell'}{2}+\fr{5}{4}\pm\fr{\ii\wt\nu}{2}\big)_{m+1}}.
\end{align}
Clearly, the rest of double series is just (\ref{eq_GBG}). 

\subsection{Equivalent expressions for general CIS}
\label{app_CIS}
Now, we can apply the above method to more general cases (\ref{eq_generalBG}), where the summand is a product of Lauricella functions $\mathrm{F}_{\mathrm{C}}$. For clarity, we explicitly assign summation indices for all dimensionless variables $r_{(\al i)}$ in these Lauricella series as follows: For an arbitrary line $i$ connecting the mother vertex $i'$ and the daughter vertex $i$, the series expansion in $r_{(ii')}$ and $r_{(ii)}$ is denoted as:
\begin{align}
  \sum_{n_i,\ob{n}_i}\cdots\times r_{(ii')}^{2n_i}r_{(ii)}^{2\ob n_i}.
\end{align}
With this notation, the range of summation of the series expansion in (\ref{eq_generalBG}) is found to be:
\begin{align}
  &\ell_i,n_i,\ob n_i\geq 0, && \ob n_i+\sum_{j\in\mathcal{I}_i/\{i\}}n_j\leq\left\lfloor \frac{\ell_i}{2}\right\rfloor.
\end{align}
Similarly, we can change the summation indices as:
\begin{align}
  &\ell_i'=\ell_i-2\ob n_i-2\sum_{j\in\mathcal{I}_i/\{i\}}n_j,&&m_i=n_i+\ob n_i.
\end{align}
Then, the original expression (\ref{eq_generalBG}) can be rewritten as:
\begin{align}
  \G_{\text{BG}}=&~2^{3V-2}\cos(\pi p_{\text{tot}}/2)\sum_{\{\ell',m\}=0}^{\infty}\Gamma(p_{\text{tot}}+\ell_{2\cdots V}'+2m_{2\cdots V}+4V-3)\n\\
  &\times\prod_{i=2}^{V}\Bigg[\frac{(-1)^{\ell'_i}}{\ell'_i!}\left(\frac{K_i}{2E_1}\right)^{2m_i+3}\left(\frac{E_i}{E_1}\right)^{\ell'_i+p_i+1}\n\\
  &\times\sum_{\ob n_i=0}^{m_i}\frac{(-1)^{\ob n_i}(p_i+\ell_i'+q_i'+2\ob n_i+5/2)}{\ob n_i!(m_i-\ob n_i)!(p_i+\ell_i'+q_i'+\ob n_i+5/2)_{m_i+1}[(p_i+\ell_i'+q_i'+2\ob n_i+5/2)^2+\wt\nu_i^2]}\Bigg].
\end{align}
Here, we redefine the family parameter $q_i'$ such that it agrees with the original definition in \cite{Liu:2024str}:
\begin{align}
  q_i'=q_i+2\sum_{j\in\mathcal{I}_i/\{i\}}n_j=\wt\ell_i'+\wt p_i+2\wt m_i+4N_i.
\end{align}
Then, using (\ref{eq_sqDiff}) and (\ref{eq_clm}), the finite summations over $\ob n_i$ can be similarly finished and give us the familiar CIS expression in \cite{Liu:2024str}.

\section{Useful functions}
\label{app_2F1}
In this appendix, we collect definitions and relevant properties of several special functions that are used throughout the main text. To set the stage, we introduce a compact notation for products and quotients of Euler $\Gamma$ functions:
\begin{align}
  \Gamma[a_1,\cdots,a_n]\equiv \Gamma(a_1)\cdots \Gamma(a_n),
\end{align}
\begin{align}
\Gamma\bigg[\bgm a_1,\cdots,a_m\\b_1,\cdots,b_n\edm\bigg]\equiv\FR{\Gamma[a_1,\cdots,a_m]}{\Gamma[b_1,\cdots,b_n]}.
\end{align}
In addition, we frequently employ the Pochhammer symbol $(a)_n$, given by:
\begin{align}
(a)_n\equiv\FR{\Gamma(a+n)}{\Gamma(a)}.
\end{align}
With these preparations, now we give the series expression of Gauss's hypergeometric function:
\begin{align}
\label{eq_2F1}
{}_2\text{F}_1\bigg[\bgm a,b\\c\edm\bigg|z\bigg]\equiv\sum_{n=0}^{\infty}\FR{(a)_n(b)_n}{(c)_n}\FR{z^n}{n!}.
\end{align}
We also use the dressed hypergeometric function in this work for simplicity, defined as:
\begin{align}
  {}_2\mathcal{F}_1\bigg[\bgm a,b\\c\edm\bigg|z\bigg]\equiv{\Gamma\bigg[\bgm a,b\\c\edm\bigg]}{}_2\text{F}_1\bigg[\bgm a,b\\c\edm\bigg|z\bigg]=\sum_{n=0}^{\infty}\Gamma\bigg[\bgm a+n,b+n\\c+n\edm\bigg]\FR{z^n}{n!}.
\end{align}
These functions are defined by series above where they converge, and by analytic continuation elsewhere. Also, we use the following transformation-of-variable identity in \cite{nist:dlmf} when studying different limits of the function $\mathcal{Z}(\wt\nu,p;r)$ defined in (\ref{eq_Z}):
\begin{align}
  {}_2\text{F}_1\bigg[\bgm a,b\\c\edm\bigg|z\bigg]=&~\Gamma\bigg[\bgm c,c-a-b\\c-a,c-b\edm\bigg]\, {}_2\text{F}_1\bigg[\bgm a,b\\a+b-c+1\edm\bigg|1-z\bigg]\n\\
  &+(1-z)^{c-a-b}\Gamma\bigg[\bgm c,a+b-c\\a,b\edm\bigg]\,{}_2\text{F}_1\bigg[\bgm c-a,c-b\\c-a-b+1\edm\bigg|1-z\bigg].
\end{align}
With this identity, it is straightforward to extract the singular term of $\mathcal{Z}(\wt\nu,p;r)$ around $r\to\pm 1$:
\begin{align}
  \lim_{r\to 1}\mathcal{Z}(\wt\nu,p;r)=&~\frac{2^p\sqrt{\pi}\ii\Gamma(p+2)}{\sinh(\pi\wt\nu)(1-r^2)^{p+2}}+\text{regular terms},\\
  \lim_{r\to -1}\mathcal{Z}(\wt\nu,p;r)=&~\lim_{r\to 1}\FR12\left[\mathcal{Z}(\wt\nu,p;re^{\ii\pi})+\mathcal{Z}(\wt\nu,p;re^{-\ii\pi})\right]\n\\
  &=\ii\sinh(\pi\wt\nu)\lim_{r\to 1}\mathcal{Z}(\wt\nu,p;r).
\end{align}

\paragraph{Associated Legendre functions}
For $x>1$, we define the associated Legendre functions of the first and second kinds as\footnote{Our convention for $\mathrm{Q}_{\be}^{\al}(x)$ differs from the conventional associated Legendre function $Q_{\be}^{\al}(x)$ in \cite{nist:dlmf} by an overall phase, $\mathrm{Q}_{\be}^{\al}(x)=e^{-\ii\pi\al}Q_{\be}^{\al}(x)$.}:
\begin{align}
  \mathrm{P}_{\be}^{\al}(x)\equiv&~\FR{\sqrt{\pi}(x^2-1)^{-\al/2}}{\cos(\pi\be)}\Bigg\{\FR{2^\be x^{\be+\al}}{\Gamma(\be-\al+1)\Gamma(1/2-\be)}{}_2\mathrm{F}_1\left[\bgm-\frac{\be+\al}{2},\frac{1-\be-\al}{2}\\1/2-\be\edm\middle|x^{-2}\right]\n\\
  &-\FR{x^{-\be+\al-1}}{2^{\be+1}\Gamma(-\be-\al)\Gamma(\be+3/2)}{}_2\mathrm{F}_1\left[\bgm\frac{\be-\al}{2}+1,\frac{\be-\al+1}{2}\\\be+3/2\edm\middle|x^{-2}\right]\Bigg\},\\
  \mathrm{Q}_{\be}^{\al}(x)\equiv&~\FR{\sqrt{\pi}\Gamma(\be+\al+1)}{2^{\be+1}\Gamma(\be+3/2)}(x^2-1)^{\al/2}x^{-\be-\al-1}{}_2\mathrm{F}_1\left[\bgm\frac{\be+\al}{2}+1,\frac{\be+\al+1}{2}\\\be+3/2\edm\middle|x^{-2}\right].
\end{align}
A useful connection formula for these functions is:
\begin{align}
\label{eq_PQconnection}
  \Gamma\left[\bgm\lambda+\alpha+1/2\\\lambda-\alpha+1/2\edm\right]\mathrm{P}_{\lambda-1/2}^{-\alpha}(x)=&~\mathrm{P}_{\lambda-1/2}^{\alpha}(x)-\FR{2\sin(\pi\alpha)}{\pi}\mathrm{Q}_{\lambda-1/2}^{\alpha}(x),\n\\
  =&~\FR{\cos[\pi(\lambda-\al)]}{\pi\sin(\pi\lambda)}\left[\mathrm{Q}_{-\lambda-1/2}^{\al}(x)-\mathrm{Q}_{\lambda-1/2}^{\al}(x)\right].
\end{align}

In the main text we have used a pair of integral representations for associated Legendre functions in (\ref{eq_Qint}) and (\ref{eq_Pint}). They can be derived from the following integral representations in \cite{nist:dlmf}:
\begin{align}
  \mathrm{Q}_{\lambda-1/2}^{\alpha}(\cosh\xi)=&~\FR{2^\alpha\sqrt{\pi}}{\Gamma(1/2-\alpha)}(\sinh\xi)^{-\alpha}\int_0^{\infty}\di u\,\FR{(\sinh u)^{-2\alpha}}{[\cosh\xi+\sinh\xi\cosh u]^{\lambda-\alpha+1/2}},\\
  \mathrm{P}_{\lambda-1/2}^{\alpha}(\cosh\xi)=&~\FR{\sqrt{2}(\sinh\xi)^\alpha}{\sqrt{\pi}\Gamma(1/2-\alpha)}\int_0^\xi\di u\,\FR{\cosh(\lambda u)}{[\cosh\xi-\cosh u]^{\alpha+1/2}}.
\end{align}
To derive (\ref{eq_Qint}), we change the integration variable in the above expression from $u$ to $t$ through the substitution: 
\begin{align}
  t=-\ln\left(\FR{e^\xi}{\cosh\xi+\sinh\xi\cosh u}\right).
\end{align}
Then it is easy to find that the $t$-integral from $0$ to $\infty$ is just the representation in (\ref{eq_Qint}). As for (\ref{eq_Pint}), we first separate the integral into two parts:
\begin{align}
    \mathrm{P}_{\lambda-1/2}^{\alpha}(\cosh\xi)=&~\FR{(\sinh\xi)^\alpha}{\sqrt{2\pi}\Gamma(1/2-\alpha)}\left[\int_0^\xi\di u\,\FR{e^{\lambda u}}{(\cosh\xi-\cosh u)^{\alpha+1/2}}+\int_0^\xi\di u\,\FR{e^{-\lambda u}}{(\cosh\xi-\cosh u)^{\alpha+1/2}}\right].
\end{align}
Then after the substitution of integration variables $t=\xi-u,t=\xi+u$ respectively, two integrals can be recombined into the $t$-integral from $0$ to $2\xi$, which is just (\ref{eq_Pint}).

\paragraph{Lauricella functions}  
The Lauricella function $\mathrm{F}_{\mathrm{C}}$ is a multivariate generalization of Gauss's hypergeometric function and is defined by:
\begin{align} 
  \mathrm{F}_{\mathrm{C}}\left[\bgm a,b\\ c_1,\cdots,c_N\edm\middle|z_1,\cdots,z_N\right]\equiv  
  \sum_{n_1,\cdots,n_N=0}^\infty(a)_{n_{1\cdots N}}(b)_{n_{1\cdots N}}\prod_{i=1}^N\FR{1}{(c_i)_{n_i}}\FR{z_i^{n_i}}{n_i!}. 
\end{align} 
When one of the lower parameters approaches a non-positive integer, the function generally becomes singular and its leading singular term can be isolated as:
\begin{align} 
\label{eq_FCn} 
  \lim_{c_1\to -n}\mathrm{F}_{\mathrm{C}}\left[\bgm a,b\\ c_1,\cdots,c_N\edm\middle|z_1,\cdots,z_N\right]\sim\Gamma(c_1)\mathrm{F}_{\mathrm{C}}\left[\bgm a+n+1,b+n+1\\ n+2,c_2,\cdots,c_N\edm\middle|z_1,\cdots,z_N\right]\frac{(a)_{n+1}(b)_{n+1}z_1^{n+1}}{(n+1)!}. 
\end{align} 
We will also use its asymptotic behavior when one of the parameters becomes large, proved in \cite{Belrhali:2026act}:
\begin{align} 
\label{eq_FCasymp} 
  &\lim_{|\mu_1|\to \infty}\mathrm{F}_{\mathrm{C}}\left[\bgm \frac{p+\ii(\cc\mu)_{1\cdots N}+1}{2}+\frac{3N}{4},\frac{p+\ii(\cc\mu)_{1\cdots N}+2}{2}+\frac{3N}{4}\\1+\ii\cc_1\mu_1,\cdots,1+\ii\cc_N\mu_N\edm\Bigg|r_1^2,\cdots,r_N^2\right]\n\\ 
  \sim &(\ii\cc_1\mu_1)^{-\ii(\cc\mu)_{2\cdots N}-(N-1)/2}\left(\frac{r_1}{2}\right)^{-\ii\cc_1\mu_1}\sum_{\dd_2,\cdots,\dd_N=\pm}\mathcal{A}_{\dd_2\cdots\dd_N}e^{-\ii\cc_1\mu_1\xi(\frac{r_1}{1+(\dd r)_{2\cdots N}})}. 
\end{align}
Here $\mathcal{A}_{\dd_2\cdots\dd_N}$ is independent of $\mu_1$. 
\end{appendix}

\newpage
\bibliography{CosmoCollider} 
\bibliographystyle{utphys}

\end{document}